\documentclass[11pt, oneside]{article}
\usepackage[margin=1in]{geometry}

\usepackage{cite}
\usepackage{amsmath,amssymb,amsfonts}
\usepackage{graphicx}
\usepackage{textcomp,nicefrac}
\usepackage{booktabs}
\usepackage{textcomp}
\usepackage{subcaption}
\usepackage{caption}
\usepackage{url}
\usepackage{authblk}

\usepackage{hyperref}
\hypersetup{
    colorlinks=true,
    linkcolor=blue,
    filecolor=blue,      
    urlcolor=blue,
    citecolor=blue,
}

\title{A CubeSat Electronics System for Dual-Satellite Coordinated Soft X-Ray Polarimetry}

\author[1]{Lirong Xie}
\author[2]{Shiqiang Zhou}
\author[2]{Kai Chen}
\author[1]{Zuke Feng}
\author[1,3]{Difan Yi}
\author[2]{Ran Chen}
\author[1]{Ruinan Fan}
\author[2]{Cheng Lian}
\author[2]{Ziyi Zhang}
\author[2]{Siying Liu}
\author[2]{Dong Wang}
\author[2]{Xiangming Sun}
\author[1]{Enwei Liang}
\author[1]{Huanbo Feng}
\author[1]{Hongbang Liu}

\affil[1]{School of Physical Science and Technology, Guangxi University, Nanning 530004, China}
\affil[2]{PLAC, Key Laboratory of Quark \& Lepton Physics (MOE), Central China Normal University, Wuhan 430079, China}
\affil[3]{School of Physical Science, University of Chinese Academy of Sciences, Beijing 100049, China}

\date{\today}

\begin{document}

\maketitle

\begin{abstract}
Addressing the unique requirements for a wide field of view and rapid response in soft X-ray polarization measurements of transient sources such as gamma-ray bursts, this paper proposes the design of a high-reliability CubeSat payload electronics system for dual-satellite cooperative observation. The system employs a Gas Microchannel Pixel Detector (GMPD) and a Topmetal-L sensor as its core components, forming a highly integrated, low-noise payload hardware platform. It achieves a field of view of 90$^\circ$ $\times$ 90$^\circ$, a sensitive area of 3.69 cm$^{2}$, and a power consumption of less than 6 W, operating within an energy range of 2--10 keV. The system incorporates autonomous high-voltage (HV) ramp-up/ramp-down control and a dual-protection mechanism based on count rate and discharge events, enabling in-orbit responses to risks such as the South Atlantic Anomaly and solar particle events. It also supports single-event upset detection and recovery, as well as in-orbit firmware upgrades. The communication interface adopts a redundant primary/backup Controller Area Network (CAN) bus design, with measured channel switching times of less than 100 ms, meeting the demands for real-time command interaction in dual-satellite coordination. By utilizing a large-array pixel sensor with region-of-interest readout and integrating an in-orbit track compression algorithm, the system significantly reduces data storage and downlink transmission resource burdens. Ground tests demonstrate an equivalent noise charge of 22.35 e$^{-}$, HV monitoring linearity better than $\pm$0.5\%, and an output range extending to -5 kV. Thermal vacuum cycling tests show no performance degradation after five cycles between -5 $^\circ$C and 40 $^\circ$C. This work demonstrates the system's capability for autonomous observation, intelligent coordination, and reliable operation in complex space environments.
\end{abstract}

\begin{center}
\end{center}

\section{Introduction}
\label{sec:1}

Since the beginning of the 21st century, breakthroughs in polarimetry have provided astrophysical research with a new and crucial dimension. The technique, first demonstrated in the 1970s by the OSO-8\cite{ref1} satellite, experienced relatively slow development until the introduction of the Gas Pixel Detector\cite{ref2,ref3} in 2001, which marked a fundamental advance in photoelectric polarimetry. This method derives polarization information from reconstructed photoelectron tracks, offering advantages such as high sensitivity, low background, and good energy resolution. Its development subsequently propelled a series of space-borne experiments, including PolarLight\cite{ref4}, IXPE\cite{ref5}, and eXTP\cite{ref6}, gradually maturing the field of soft X-ray polarimetry.

Gamma-Ray Bursts (GRBs)\cite{ref7} are the most energetic explosive phenomena in the universe since the Big Bang, serving as natural laboratories for studying extreme astrophysics and relativistic jets. Polarimetric measurements of GRBs are regarded as a key probe for uncovering the central engine mechanism, jet structure, magnetic field configuration, and particle acceleration processes\cite{ref8,ref9,ref10,ref11}.

However, capturing and measuring such signals imposes stringent demands on a detection system. It requires a wide field of view and rapid response to capture transient events, while delivering high sensitivity, low noise, and excellent energy and spatial resolution to measure faint signals precisely\cite{ref12}. The large data volume from photoelectron track imaging further challenges the limited downlink bandwidth, necessitating onboard real-time processing and compression. 

Traditional space-based observations have relied on large, single satellites, which offer high performance but suffer from long development cycles, high cost, limited field of view, and limited temporal coverage. The advancement of micro- and nano-satellite technologies, alongside maturing constellation architectures, is enabling a new paradigm of distributed, cooperative observation\cite{ref13}. Networks of low-cost, standardized small satellites can achieve breakthroughs in spatial coverage, multi-perspective observation, and system redundancy — capabilities especially valuable for monitoring transient sources like GRBs. The CubeSat, with its standardization, modularity, and rapid deployment potential, has emerged as an ideal candidate for building such networks.

Nevertheless, CubeSats operate under stringent constraints in volume, mass, power, thermal management, and communication bandwidth\cite{ref14}. These limitations impose demanding requirements on payload design, necessitating high integration, power efficiency, robustness, and intelligent onboard processing. Effective multi-satellite coordination further introduces challenges in inter-satellite communication, attitude synchronization, and mission planning\cite{ref15}. Consequently, the deep integration of capable scientific payloads with the satellite bus, coupled with embedded autonomous and cooperative intelligence, has become a critical research direction in modern space science.

Against this backdrop, the CXPD (Cosmic X-ray Polarization Detection) CubeSat mission\cite{ref16}, serving as a prototype for the future POLAR-2/LPD\cite{ref17,ref18,ref19} payload scheduled for the China Space Station, has successfully achieved wide-field soft X-ray polarimetric detection in orbit. The mission completed polarization observations of two bright X-ray sources, Sco X-1 and Swift J1727.8-1613, obtaining preliminary constraints on their polarization degree and angle\cite{ref20}. The mission has accumulated crucial technical and engineering experience to support the subsequent deployment of a polarimeter with a wider field of view and enhanced sensitivity on the space station. 

To address these scientific demands and technical challenges, this paper proposes a CubeSat payload electronics system for dual-satellite cooperative observation. The CXPD-03 and CXPD-04 payloads (hereinafter jointly referred to as the CXPD-Duo payload configuration) represent a systematic upgrade based on the existing CXPD-01 technology, centered around three core objectives: cooperative observation, autonomous processing, and reliable operation. The structure of this paper is as follows: Section~\ref{sec:2} defines the dual-satellite cooperative requirements, focusing on the key data flows and design challenges for the payload electronics. Section~\ref{sec:3} elaborates on the hardware system design of the payload. Section~\ref{sec:4} introduces the implementation of the onboard intelligent processing firmware. Section~\ref{sec:5} presents the results of ground testing and verification. Section~\ref{sec:6} provides a conclusion.

\section{System Requirements for CXPD-Duo}
\label{sec:2}

\begin{figure}[t]
\centerline{\includegraphics[width=3.5in]{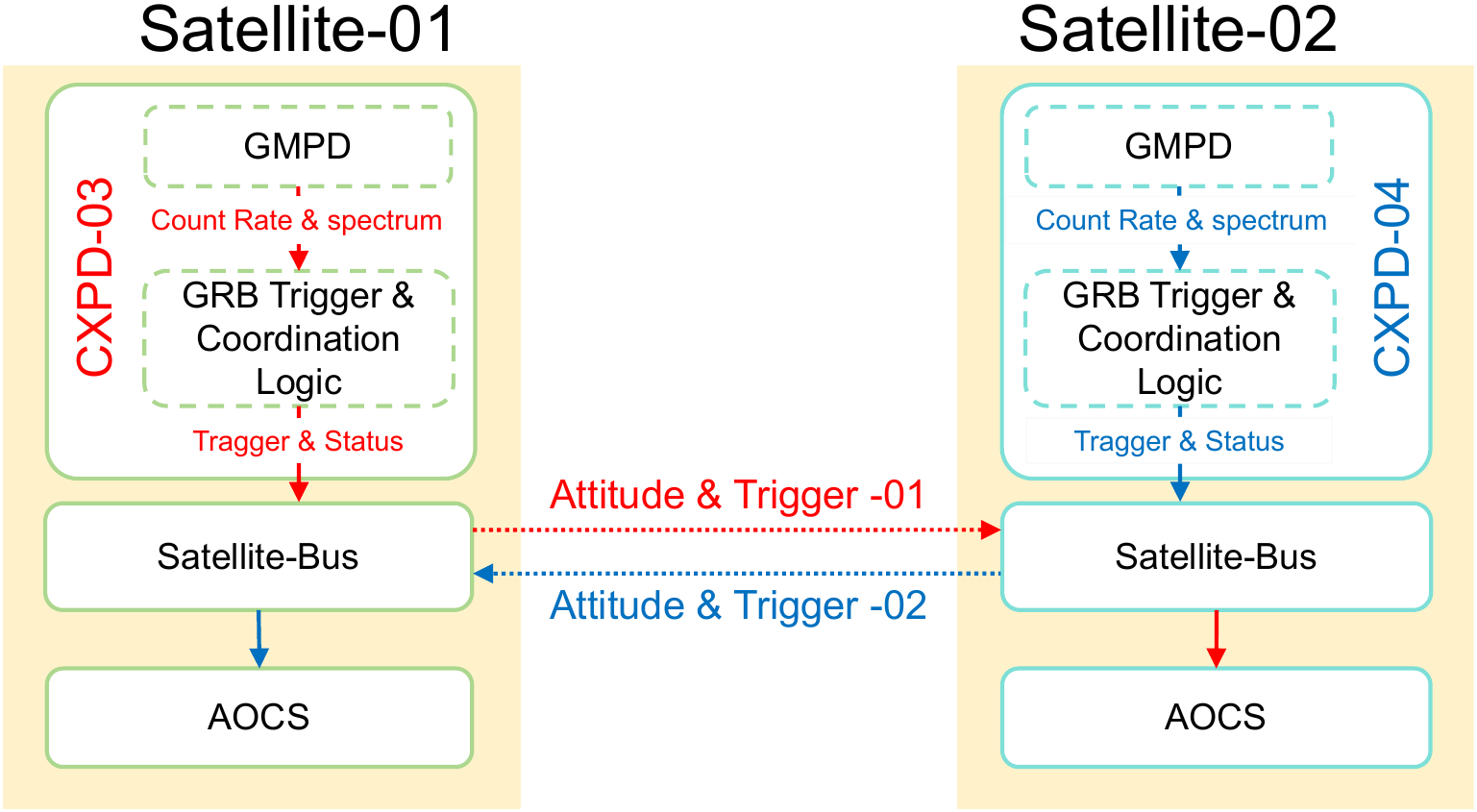}}
\caption{Dual-satellite cooperative signal flow. CXPD-03 and CXPD-04 are mounted on Satellite-01 and Satellite-02, respectively. Red arrows indicate the flow when CXPD-03 detects a GRB; blue arrows for CXPD-04.}
\label{fig:1}
\end{figure}

\begin{figure}[t]
\centerline{\includegraphics[width=2.6in]{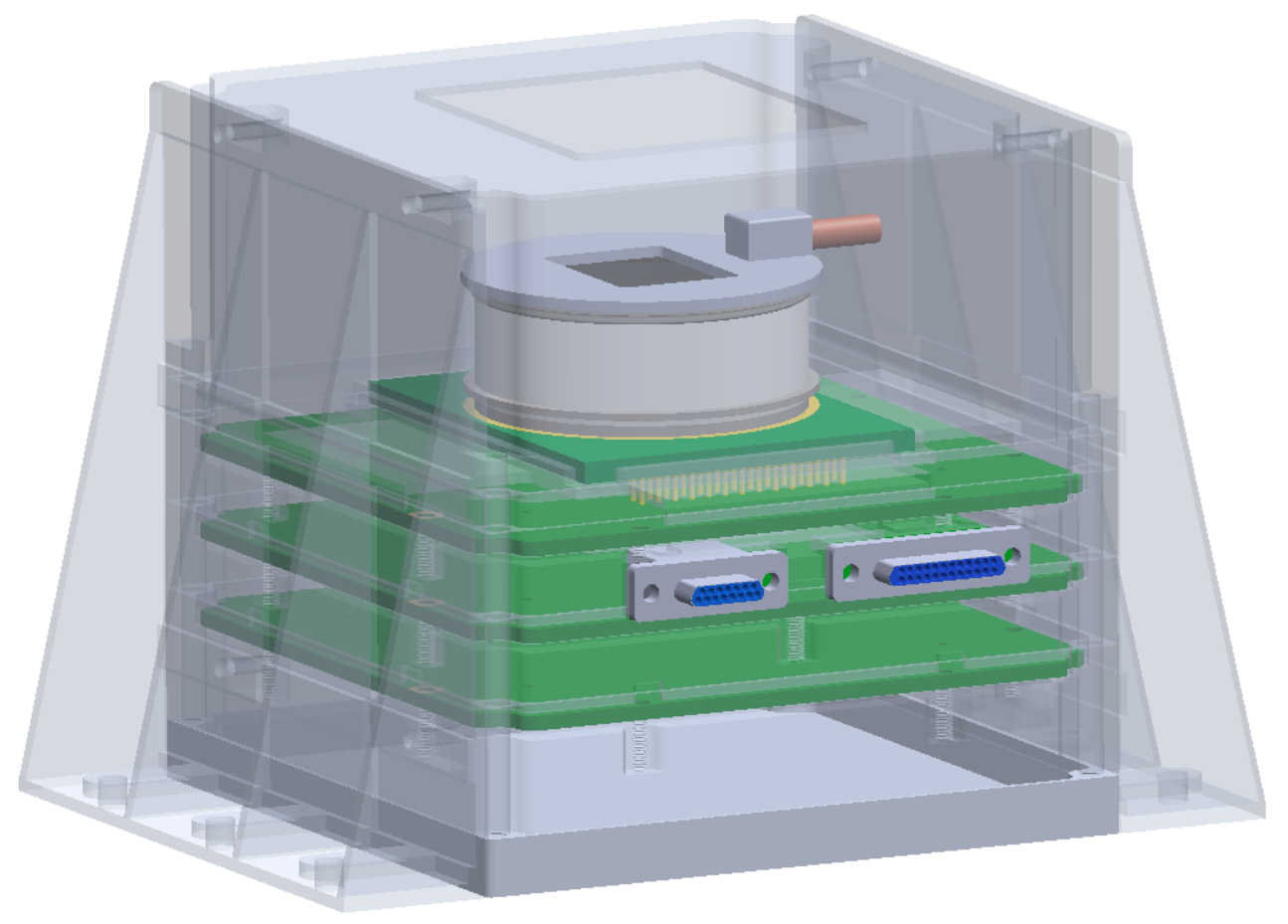}}
\caption{The mechanical structure of CXPD-Duo.}
\label{fig:2}
\end{figure}

\begin{figure}[t]
\centerline{\includegraphics[width=3in]{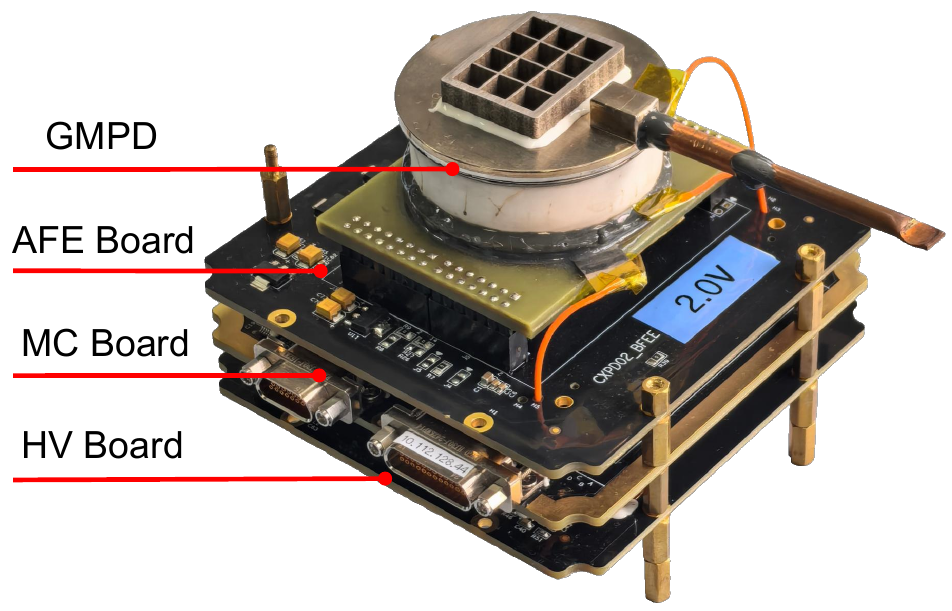}}
\caption{Integrated physical view of the CubeSat payload electronics system with the GMPD. The required high voltage for the GMPD is supplied via a dedicated HV flying wire connected to the corresponding electrode.}
\label{fig:3}
\end{figure}

\begin{figure*}[t]
\centerline{\includegraphics[width=5.38in]{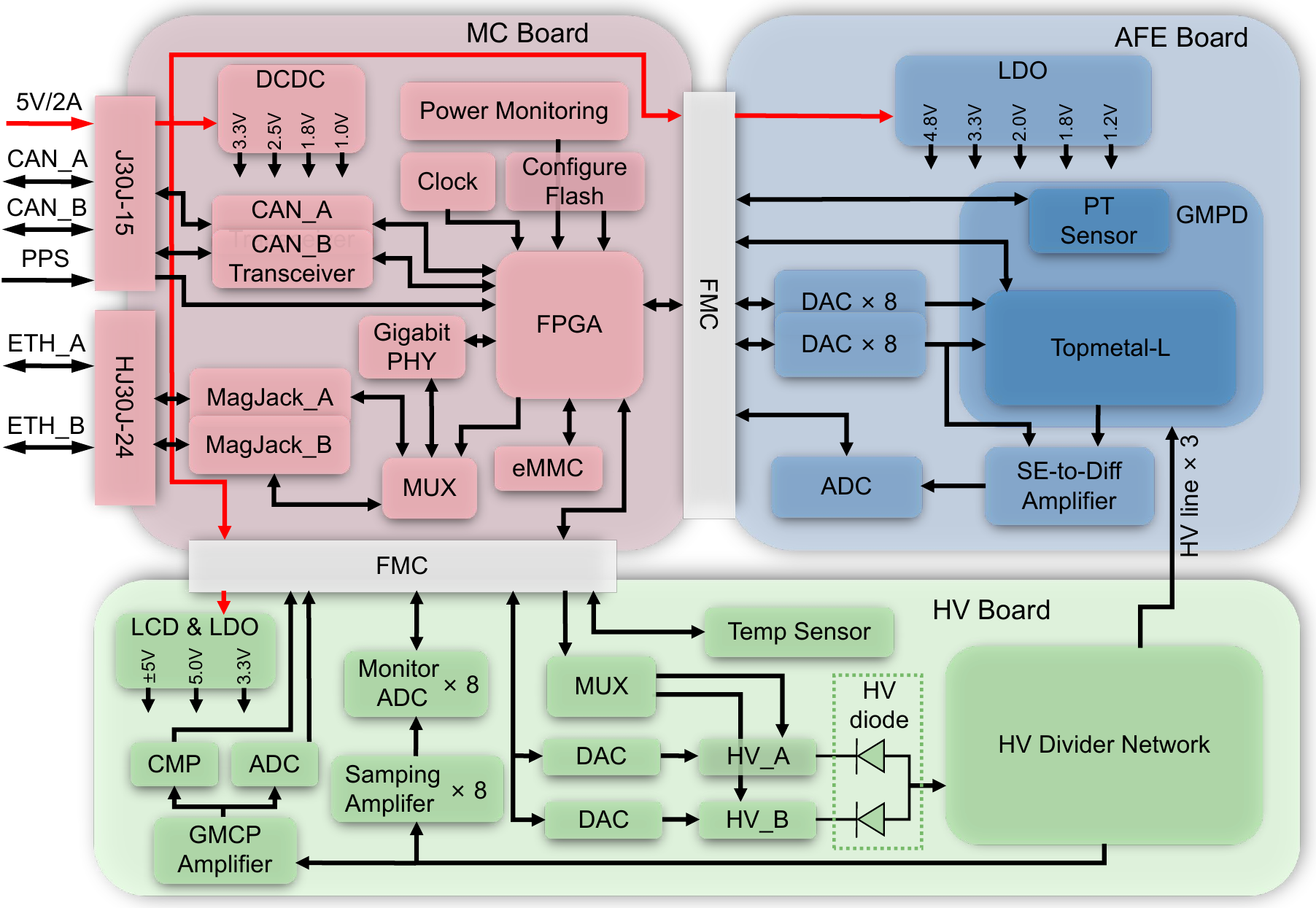}}
\caption{Schematic of functional modules and external interfaces for each electronics board.}
\label{fig:4}
\end{figure*}

To validate the multi-satellite cooperative observation paradigm, the CXPD-Duo payload configuration establishes an in-orbit verification system. As illustrated by the signal flows in Fig.~\ref{fig:1}, red arrows indicate the flow when CXPD-03 acts as the primary satellite, and blue arrows indicate the flow when CXPD-04 acts as the primary. The end-to-end workflow is: the detecting satellite’s trigger algorithm issues a collaboration flag while the Satellite Integrated Avionics (SIA) provides real-time navigation data;this combined packet is linked to the companion; the companion then verifies its own status, commands its Attitude and Orbit Control System (AOCS) for coordinated pointing, and finally downlinks the joint science data. The implementation of this complete chain necessitates an evolution of payload electronics beyond traditional architectures. The required capabilities can be divided into baseline payload requirements and dual-satellite cooperative requirements.

\textbf{Baseline payload requirements:}

\begin{itemize}
    \item \textbf{High-Voltage Safety Management:} An autonomous high-voltage (HV) manager employs zone-specific logic and dual-protection mechanisms against discharge events and excessive count rates to ensure the safety of the Gas Microchannel Pixel Detector (GMPD)\cite{ref21}.
    
    \item \textbf{Low-Noise Readout:} The fidelity of soft X-ray polarization measurement depends on an ultra-low-noise analog front-end to preserve the minute structure of photoelectron tracks.
    
    \item \textbf{Data Compression:} Real-time compression of the pixel sensor image data is implemented to alleviate downlink bandwidth constraints.
\end{itemize}

\textbf{Dual-satellite cooperative requirements:}

\begin{itemize}
    \item \textbf{Onboard Real-Time Autonomous Processing:} This onboard intelligence enables either satellite, upon detecting a GRB, to act as the primary and issue a valid trigger, while the other satellite autonomously verifies its readiness for a coordinated slew.
    
    \item \textbf{Communication Reliability:} Module-level redundancy and firmware hardening ensure reliable control and communication in the space environment, which is fundamental for robust joint decision-making between the payload and the platform.
\end{itemize}

The CXPD-Duo electronics serve as more than a simple data acquisition unit. They evolve into the autonomous decision-making core of the cooperative network, enabling a new class of distributed space science observations.

\section{CubeSat Payload Hardware System}
\label{sec:3}
\subsection{System Architecture and Hardware Integration}

The core function of the CXPD series payload electronics system is to supply the GMPD with the required low-voltage power, HV bias, analog bias, signal readout, and digital control, ensuring the detector operation. Designed to observe cosmic X-ray sources and the soft X-ray components of GRBs, the system is deployed on a near-Earth sun-synchronous orbit satellite bus. This orbit was chosen for its cost‑effective rideshare launch opportunities, consistent with the pathfinder nature of the mission. The satellite’s downlink is used for data transmission and subsequent ground analysis. Mission control and health monitoring are performed indirectly through the SIA. Additionally, the bus is equipped with onboard intelligent processing capabilities based on a multimodal large language model\cite{ref22}, which enables the implementation of an autonomous GRB identification strategy specifically designed for the CXPD payload.

The mechanical structure of CXPD-Duo, shown in Fig.~\ref{fig:2}, comprises an aluminum-alloy main body coated with white paint. The payload is built around a 96 $\times$ 96 $\times$ 96 mm$^3$ cubic core, with two triangular prism brackets attached to its sides as mechanical interfaces coordinated with the satellite platform. These brackets are mounted to the satellite via six base mounting holes, providing both mechanical support and a thermal path to the satellite bus. The GMPD detector and electronics boards are fixed to the aluminum structure using mounting holes, ensuring both mechanical stability and a thermal path. To ensure operational safety under HV bias and prevent plasma-induced discharges, the top X-ray entrance window is sealed with an aluminum-coated polyimide film. Selected design parameters of CXPD-Duo are listed in Table~\ref{tab:1}.

\begin{table}[!htp]
\caption{}
\label{tab:1}
\scshape
\centering
\small
Technical Specifications of the CXPD-Duo Payload
\normalfont
\vspace{8pt}

\centering
\begin{tabular}{p{6cm} p{4cm}}
\toprule
\textbf{Parameter} & \textbf{Specification} \\
\midrule
Overall Dimensions & $96 \times 140 \times 96\,\text{mm}^3$ \\
Surface Roughness & $3.2\,\mu\text{m}$ \\
Mass & $760\,\text{g}$ \\
Total Power Consumption & $\leq 6\,\text{W}$ \\
Energy Range & $2$--$10\,\text{keV}$ \\
Sensitive Area & $3.69\,\text{cm}^2$ \\
Field of View & $90^\circ \times 90^\circ$ \\
Energy Resolution & $\leq 25\% \,@ \,5.9\,\text{keV}$ \\
Polarization Modulation Factor & $\geq 40\% \,@ \,5.9\,\text{keV}$ \\
\bottomrule
\end{tabular}


\end{table}

Fig.~\ref{fig:3} illustrates the physical layout of the integrated CXPD-Duo electronics system. In the circuit design, several measures were implemented to meet electromagnetic compatibility requirements, including low-impedance grounding, bonding at critical nodes, partitioned routing, and signal isolation. The layout of power lines and low-frequency signal traces strictly adheres to the principle of minimizing return-path loops to reduce interference. The GMPD detector, responsible for soft X-ray polarimetry, is positioned at the top of the system and mounted directly on the first circuit board in the electronics stack. The electronics system consists of three circuit boards: the Analog Front-End (AFE) board handles GMPD configuration and signal readout; the Main Controller (MC) board performs data processing, storage, system control, power management, and communication functions; and the HV board provides HV configuration, monitoring, and an analog readout channel. 

The functional modules and external interfaces of each board are shown in Fig.~\ref{fig:4}, where blue, red, and green regions correspond to the AFE, MC, and HV boards, respectively. The MC board connects to the AFE and HV boards via FPGA Mezzanine Card (FMC) interfaces. The HV board generates the HV bias required by the GMPD; this bias is routed through flying wires that pass through the MC and AFE boards to the corresponding detector electrodes. The following subsections provide a detailed description of the design of each hardware module.

\subsection{GMPD and Topmetal-L Readout}

\begin{figure}[t]
\centerline{\includegraphics[width=2.6in]{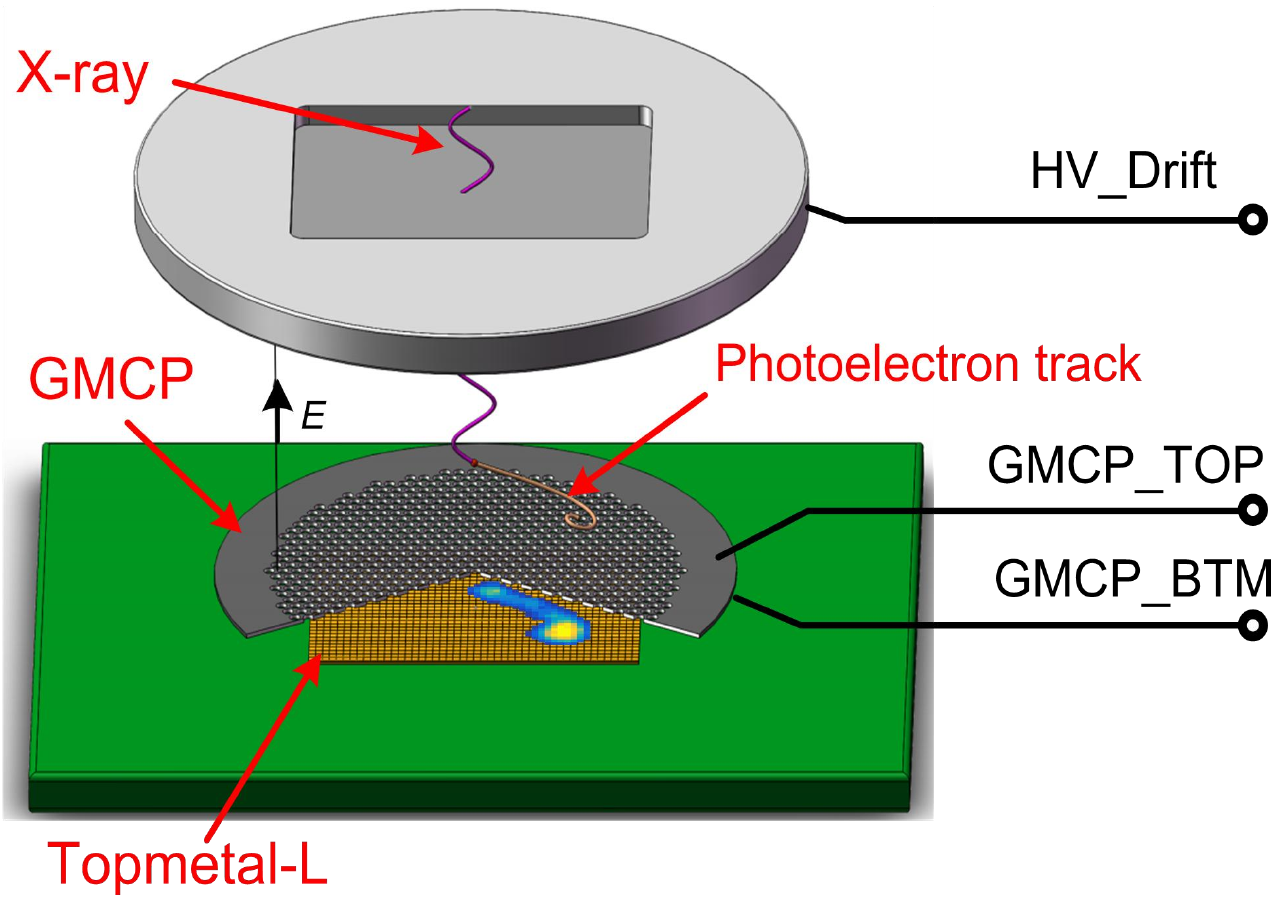}}
\caption{Schematic of GMPD operating principle.}
\label{fig:5}
\end{figure}

\begin{figure}[t]
\centerline{\includegraphics[width=3.5in]{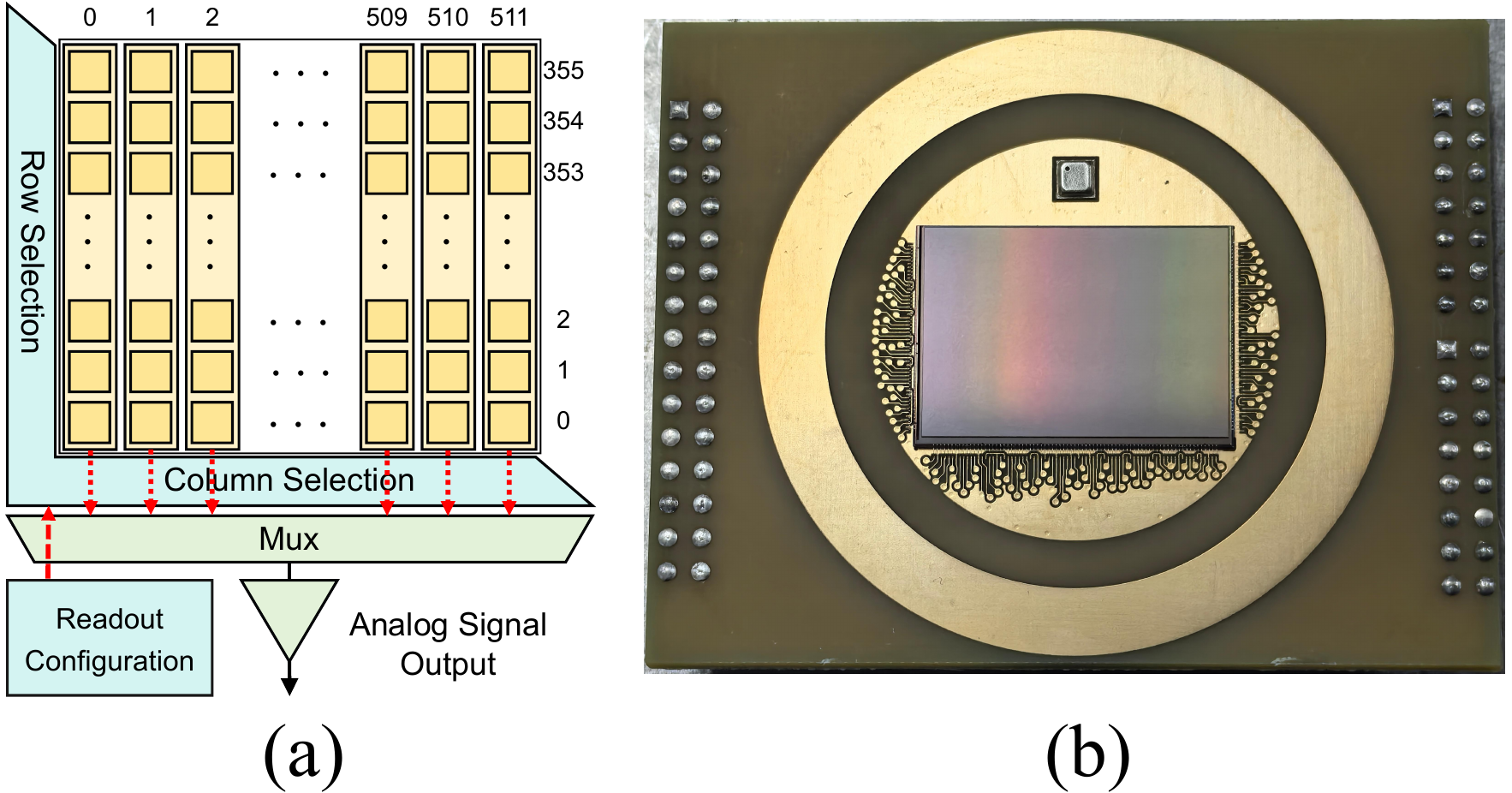}}
\caption{Topmetal-L pixel sensor. (a) Schematic diagram of the sensor architecture; (b) Photograph of the GMPD substrate, showing the Topmetal-L chip at the center, a temperature-pressure sensor above it.}
\label{fig:6}
\end{figure}

The core components of the GMPD are the Gas Microchannel Plate (GMCP)\cite{ref23} and the Topmetal pixel sensor, both housed within a sealed chamber filled with a working gas mixture. The chamber maintains a gas pressure of approximately 0.8 atmospheres, consisting of a mixture of 40\% helium and 60\% dimethyl ether. As illustrated in Fig.~\ref{fig:5}, the operational process is as follows: X-rays enter through a thin beryllium window at the top of the chamber and undergo photoelectric interaction within the gas. The resulting photoelectrons are emitted in a direction modulated by the polarization of the incident photons. These photoelectrons travel through the gas, ionizing atoms along their paths and generating electron-ion pairs. These electrons drift toward the GMCP under the influence of the applied electric field. Both surfaces of the GMCP are coated with conductive layers. When a voltage is applied, a strong electric field is established within its microchannels, enabling electron multiplication. Electrons entering these high-field regions undergo avalanche multiplication. The secondary electrons generated continue moving downward; a portion of the secondary electrons are collected by the bottom electrode of the GMCP, while the remainder are collected by the pixel sensor acting as the anode. Each pixel incorporates a charge-sensitive amplifier that converts the collected charge into a voltage pulse. The system records the position and energy information of all pixels for each X-ray event. By accumulating a large number of such events, a two-dimensional track image can be reconstructed. Subsequent morphological analysis and statistical fitting of these images ultimately yield the polarization information of the incident X-rays\cite{ref24}.

The pixel sensor and its base structure within the GMPD are shown in Fig.~\ref{fig:6}. The assembly includes a Topmetal-L\cite{ref25} and an integrated temperature-pressure sensor. The Topmetal-L features an active area of 16 $\times$ 23 mm$^{2}$ with a 356 $\times$ 512 pixel array and a 45 $\mu$m pixel pitch, sufficient spatial resolution to resolve photoelectron track structures in the 2--10 keV band. It dissipates 720 mW and supports region-of-interest readout\cite{ref26}, selectively reading only triggered pixels to reduce onboard data volume. Compared to the Topmetal-II$^{-}$\cite{ref27} used in CXPD-01, the Topmetal-L offers improved power efficiency, readout speed, and area utilization, better suiting the wide-field transient survey requirements of CubeSat missions. A temperature-pressure sensor is integrated into the sealed chamber to monitor gas integrity in real time, essential for stable detector operation given the strong dependence of photoelectric and multiplication processes on the gas environment. The sensor operates at a sampling rate of 1 Hz, achieving an absolute pressure accuracy of $\pm0.6$ hPa and an absolute temperature accuracy of $\pm1\,\text{\textcelsius}$.

The GMPD interfaces with the AFE board using two keyed dual-row connectors, which separate analog and digital signal paths to minimize digital interference in the analog front-end. The functional module layout of the AFE board is highlighted in blue in Fig.~\ref{fig:4}. Its design centers on the Topmetal-L sensor: two 8-channel, 12-bit DACs bias the sensor and subsequent amplifiers. The single-ended analog output from the sensor is converted to a differential signal by an amplifier stage and then digitized by a 12-bit ADC with a 2 V input range and a 40 MSPS sampling rate.

\subsection{HV Supply and Monitoring Circuit}

\begin{figure}[t]
\centerline{\includegraphics[width=2.9in]{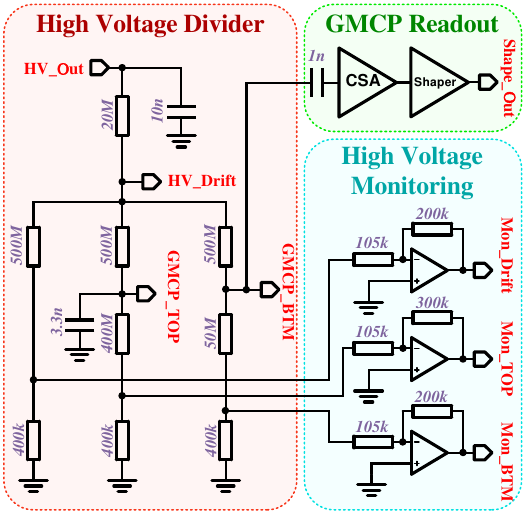}}
\caption{Schematic of the HV divider network, monitoring circuit, and GMCP bottom electrode readout circuit.}
\label{fig:7}
\end{figure}

\begin{figure*}[t]
\centerline{\includegraphics[width=5.36in]{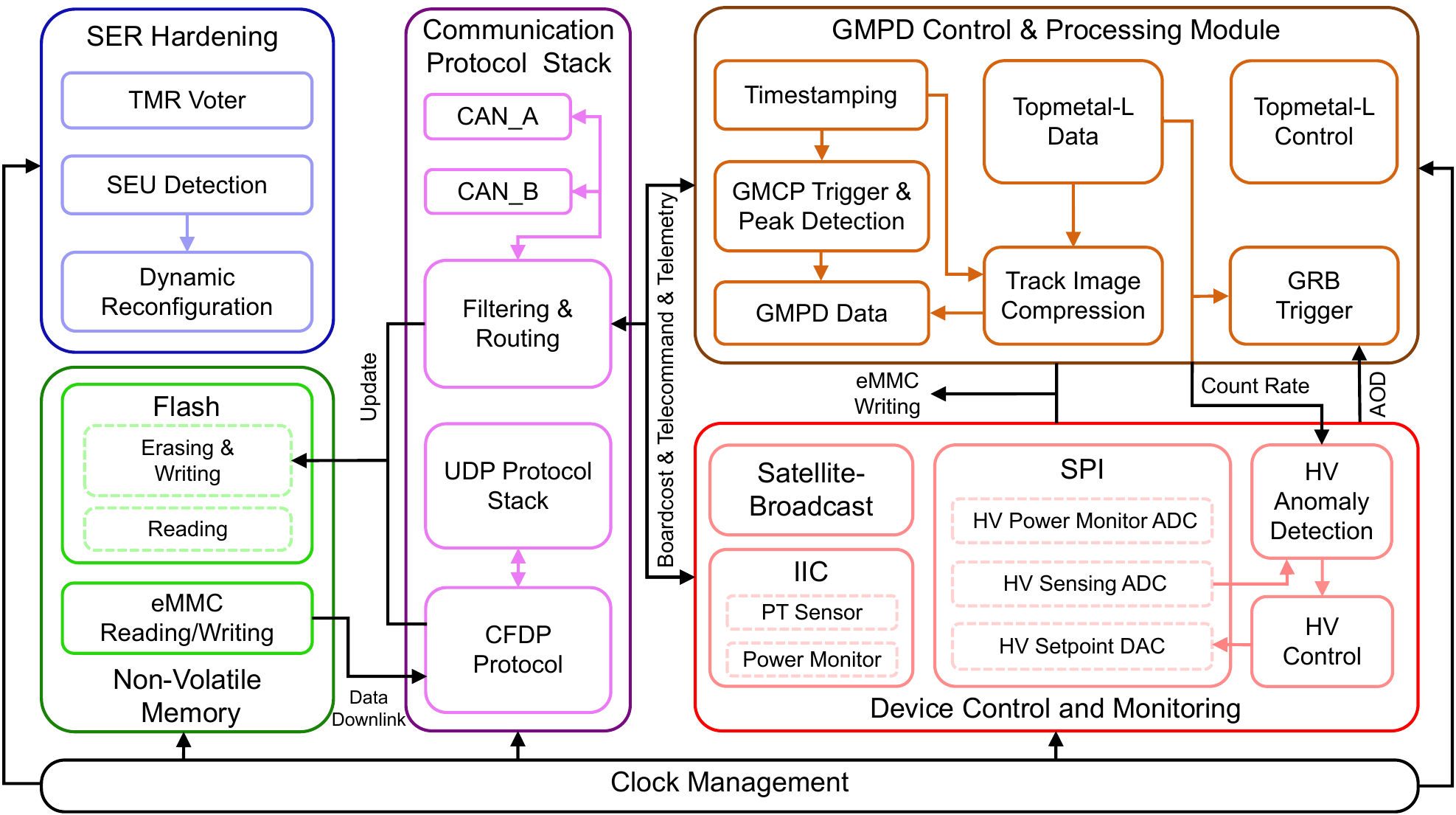}}
\caption{Block diagram of FPGA firmware functional modules.}
\label{fig:8}
\end{figure*}

The HV supply is a critical requirement for the proper operation of the detector. The strong electric field inside the GMCP is generated by the HV module. While high voltage is essential for detection performance, it also constitutes a significant risk factor in the electronics system. To address this, the HV board of CXPD-Duo has been specifically optimized for HV monitoring and long-term stability.

The HV output module employs a cold-redundant architecture. Switching between redundant channels is controlled via MOSFET power switches, with HV diodes providing electrical isolation at the output stage. HV conversion is accomplished using the compact DC-to-negative-high-voltage device UMHV0550N, which provides a maximum output current of 100 $\mu$A and can deliver negative voltages up to -5000 V. Its output voltage is set by a 12-bit rail-to-rail DAC through its control pin. The operating HV required by the GMPD is distributed through a resistive divider network. As shown in Fig.~\ref{fig:7}, each branch terminates at a sampling resistor. Together with inverting amplifier stages and an ADC, these branches form closed-loop monitoring circuits. Monitoring is performed by an 8-channel, 12-bit ADC with a maximum sampling rate of 200 kSPS per channel. In this design, the ADC operates in an 8-channel cyclic sampling mode, achieving a per-channel sampling rate of 10 kSPS. The gain of each inverting amplifier is adjusted according to the static current of its corresponding branch to match the ADC’s input voltage range. This ADC is also used to monitor the input current and supply voltage of the HV module. The ADC self-test can be initiated by an external command; upon reception, the firmware selects an internal reference voltage via SPI and compares the conversion result with the expected value, issuing a telemetry warning if a persistent deviation is detected.

A readout circuit for signals from the GMCP bottom electrode is also integrated on the HV board. A branch is taken from the GMCP\_BTM node and AC-coupled through an HV capacitor to extract weak charge signals from the charge collected at the GMCP bottom electrode\cite{ref28}. After amplification and shaping, this signal is used for timestamping and auxiliary energy measurement. Together with the Topmetal-L sensor, this readout channel forms a dual-path data acquisition module for the detector. The two signal paths are matched using time-coincidence techniques, enabling reliable event detection while effectively rejecting false triggers.

The HV board further incorporates a temperature sensor with a sampling rate of 1 Hz and an accuracy of  $\pm0.2\,\text{\textcelsius}$ for monitoring local board temperature. This sensor, together with the temperature-pressure sensor inside the GMPD chamber and the temperature acquisition channels of the FPGA’s internal XADC, constitutes a comprehensive payload temperature monitoring system for the CXPD-Duo.

\subsection{Storage, Communication and Interfaces}

The MC board serves as the central control unit for the X-ray polarimetry payload. Its core functions include detector data acquisition and management, control of various electronic modules within the payload, and telecommand, telemetry and data-file transfer with the satellite bus. The board integrates key components such as main FPGA, eMMC storage, Flash memory, primary and backup Controller Area Network (CAN) transceivers, an Ethernet module, and power-monitoring circuits. It provides two external interfaces: an HJ30J-24 dual-gigabit Ethernet connector for high-speed data transmission between the payload and the satellite bus, and a J30J-15 connector providing base services, including power, telecommand, telemetry, and the Pulse Per Second (PPS) signal. The satellite-to-ground downlink is handled by the platform’s Ka-band transmitter at a nominal rate of 370 Mbps. 

For data storage, a 64 GB eMMC compliant with the eMMC 5.0 specification is employed, supporting block management and wear-leveling. To ensure reliable data storage, the interface adopts the eMMC standard high-speed mode and operates at 52 MB/s. The system firmware resides in a 1 Gb BPI-16 parallel Flash memory. This interface delivers higher parallel read/write bandwidth, significantly reducing file transfer time. This meets the stringent timing constraints of ground-contact windows, improving the success rate and reliability of remote reconfiguration. It thereby supports dynamic in-orbit updates to dual-satellite coordination algorithms and control logic. Both the eMMC and Flash memory feature hardware-based Error-Correcting Code (ECC) mechanisms. These automatically detect and correct single-bit errors through appended parity bits, significantly enhancing memory resilience against Single-Event Effects (SEE).

In addition, the telecommand, telemetry and data channels are designed with hardware redundancy. Primary and backup CAN transceivers, together with the Ethernet module, form the external communication interface. The CAN bus employs a differential serial communication scheme, offering low noise and strong resistance to electromagnetic interference. In the space environment, it is particularly suitable for transmitting high-reliability information such as telecommands, telemetry data, time-synchronization messages, and orbit and attitude broadcasts. The Ethernet module complies with triple-speed Ethernet physical-layer standards and supports auto-negotiation and polarity correction, ensuring broad compatibility. Its physical-layer signals are routed through an Ethernet switch to two independent network transformers, enabling redundancy and multiplexing of the physical channels.

\section{Information Processing and Autonomous Management Architecture}
\label{sec:4}
\subsection{Firmware System Architecture}

The FPGA firmware functional modules and their interconnections are illustrated in Fig.~\ref{fig:8}. The purple section represents the communication and protocol module, handling interaction among the FPGA, the SIA, and the onboard router. This module filters and distributes commands and data received via the CAN bus, identifying instructions targeted for CXPD-Duo and routing them to the appropriate functional modules based on flag bits embedded in the command structure. Data transmission utilizes the CCSDS (Consultative Committee for Space Data Systems) File Delivery Protocol (CFDP)\cite{ref29}. This design makes CFDP particularly suitable for the high-latency, intermittently connected environments typical of space missions. The implemented CFDP module supports bidirectional operation, handling both scientific data downlink and onboard reception of firmware updates.

Device control and monitoring include power supply monitoring, internal temperature-pressure monitoring, HV monitoring, and integration of satellite-broadcast data, including time, orbit, and attitude information.

The GMPD control and processing module is responsible for reading out and controlling the Topmetal-L sensor and the GMCP bottom electrode. It serves as the primary source of observational data, accounting for over 80\% of the total data volume. Its core processing flow includes: timestamping the image data from Topmetal-L and applying image compression; performing trigger recognition and peak extraction on the GMCP bottom electrode readout signal, along with timestamping. Concurrently, this module performs real-time event counting for GMPD-detected events, providing inputs for GRB trigger decisions.

The green section represents the non-volatile memory module. CXPD-Duo supports in-orbit remote firmware updates. This approach stores two complete functional images in the FPGA’s firmware memory: a Golden Image (GI) and an Update Image (UI). During normal startup, the system preferentially loads the UI. When an in-orbit upgrade is required, an update file is uploaded from the ground and reprograms the UI. The GI is immutable via the update process, ensuring that the system can automatically revert to it if the UI fails to load, thereby preserving basic functionality and operational stability. During observation, the scientific data from the GMPD, system monitoring data, and satellite-broadcast data are first buffered in the eMMC before being downlinked during suitable ground-contact windows.

CXPD-Duo operates on a low-Earth sun-synchronous orbit that traverses the inner radiation belt and the South Atlantic Anomaly (SAA)\cite{ref30}. To mitigate SEE, a dual-layer hardening strategy is implemented, corresponding to the blue section in Fig.~\ref{fig:8}. The first layer employs Triple-Modular Redundancy (TMR) for critical logic modules such as HV control, HV anomaly detection and GRB trigger logic. A majority-voting mechanism masks transient errors caused by Single-Event Upsets (SEU). The system implements a second-layer defense using the FPGA's Soft Error Mitigation (SEM) IP core for dynamic SEU detection and repair\cite{ref31}. The SEM core provides real-time error counts and status. If errors exceed a threshold or become unrecoverable, a graded protection sequence is triggered: HV is disabled first to prevent damage; the payload enters safe shutdown; and an alert is sent via telemetry for ground reset.

\begin{figure*}[t]
\centerline{\includegraphics[width=6.18in]{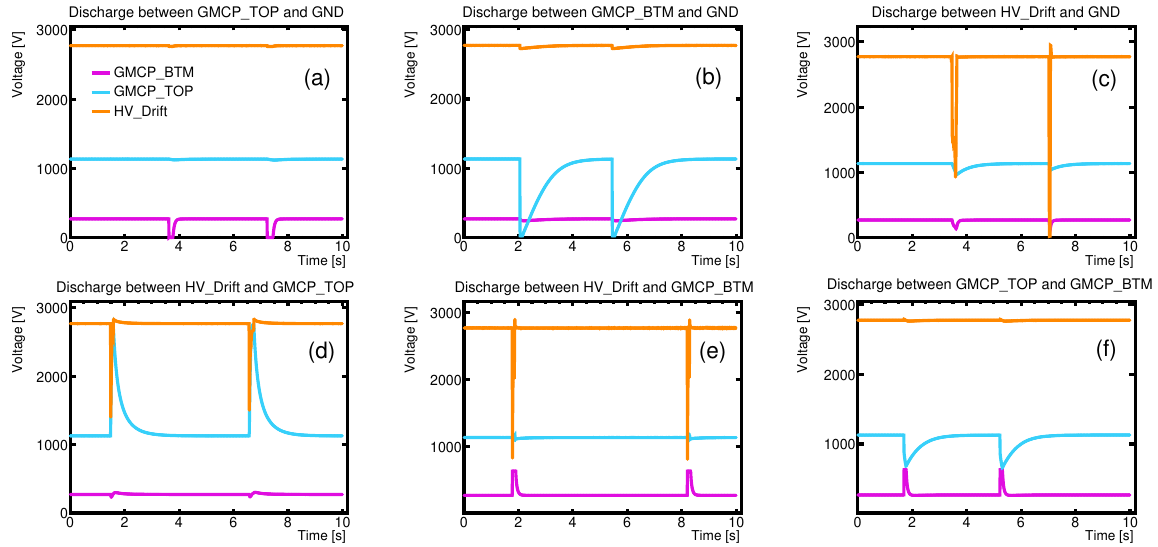}}
\caption{Typical discharge waveforms between HV electrodes. (a)--(e) Abnormal discharge events. (f) A normal discharge event. For clarity, the actual negative HV signals are displayed as their absolute values.}
\label{fig:9}
\end{figure*}

\subsection{HV Monitoring and Autonomous Control}

The GMPD requires a negative HV bias exceeding 3000 V, posing inherent safety risks. To manage this safely, the system employs a closed-loop controller with a beat timer for precise ramp-up and ramp-down control. The timer pulse frequency is dynamically adjusted based on real-time comparisons between the current voltage, a preset safe voltage, and the target voltage. Output voltage is changed by a fixed step per pulse, gradually approaching the target to ensure controlled slew rate. The target voltage can be dynamically set to a configured value, a safe level, or zero, enabling secure external control of the HV output while maintaining operational safety.

Discharge characteristics depend on electrode structure. The GMCP, with a thickness of 300 $\mu$m, operates under a potential difference exceeding 1 kV between its electrodes, making this region prone to discharge events. As shown in Fig.~\ref{fig:9} (absolute voltage), an example of such an inter-electrode discharge is presented in panel (f). Events of this type are classified as normal discharges. Their occurrence frequency correlates with operating voltage and detector count rate, and they are managed with a tolerant strategy. In contrast, waveforms in panels (a)--(e) represent abnormal discharges, which require immediate protective action to prevent damage. Abnormalities are identified using the following criteria: an abrupt drop in either HV\_Drift or GMCP\_BTM voltage triggers an abnormal-discharge response and initiates voltage reduction. A sharp drop in GMCP\_TOP voltage prompts a check of GMCP\_BTM; if no concurrent rise is observed, the event is classified as an abnormal discharge; otherwise, it is considered a normal discharge. For normal discharges, the management strategy incorporates three command-configurable parameters: Discharg\_Thr (discharge count threshold), Protect\_Thr (protection trigger threshold), and Safe\_Vol (safe voltage). These parameters are on-orbit adjustable: Discharge\_Thr and Protect\_Thr each range from 0 to 255 in steps of 1, and Safe\_Vol ranges from 0 to 3500 V with a step of about 1.22 V. The system counts discharge events over a five-minute window. If the count exceeds Discharge\_Thr, HV protection is triggered, reducing the output voltage to Safe\_Vol. If no discharge occurs for ten seconds, the voltage is restored to the operational level. If the number of protection triggers within twelve minutes exceeds Protect\_Thr, the condition is judged as an HV anomaly, and the HV output is disabled. 

For any monitored channel, the relationship between its monitored voltage and the output of the HV controller is described as follows:

\begin{equation}
\label{eq:HV_monitor}
HV_{\mathrm{MON}} - V_{\mathrm{offset}} =
\frac{2500 \, R_S \, A_S \, HV_{\mathrm{SET}}}{R_A}
\end{equation}

where $HV_\text{MON}$ is the ADC code from the HV monitoring circuit, $V_\text{offset}$ corrects the circuit's DC offset, $HV_\text{SET}$ is the DAC code driving the HV output, $R_S$ is the sampling-node-to-ground resistance, $R_A$ is the total output-to-ground resistance, and $A_S$ is the inverting-amplifier gain. The HV module delivers up to -5 kV, while the ADC input is limited to 2 V. Since all hardware parameters are constant, the relation reduces to a first-order linear formula, implemented with a fixed-point multiplier to convert the DAC code to its expected monitoring value. This expected value is compared with the actual ADC reading. A persistent large discrepancy between $HV_\text{MON}$ and $HV_\text{SET}$ indicates a possible short circuit in the voltage-divider network, also triggering an HV anomaly shutdown.

An automated HV management system supports autonomous satellite operation under space-environment constraints. When traversing SAA or during attitude maneuvers that direct the Sun into the detector’s field of view, high-energy particle and X-ray fluxes can drive the detector count rate beyond its normal range. To prevent saturation or damage, the system automatically disables the HV output before entering such regions. This logic uses real-time broadcast data from the SIA via the CAN bus. The system continuously compares this information with predefined SAA models and pointing-exclusion zones\cite{ref19,ref20}. Once conditions are met, an HV shutdown is triggered; the high voltage is restored only after the satellite exits the affected area. This automated process autonomously mitigates risks for CXPD-Duo, ensuring long-term payload safety in complex orbital environments.

\begin{figure}[t]
    \centerline{\includegraphics[width=3.5in]{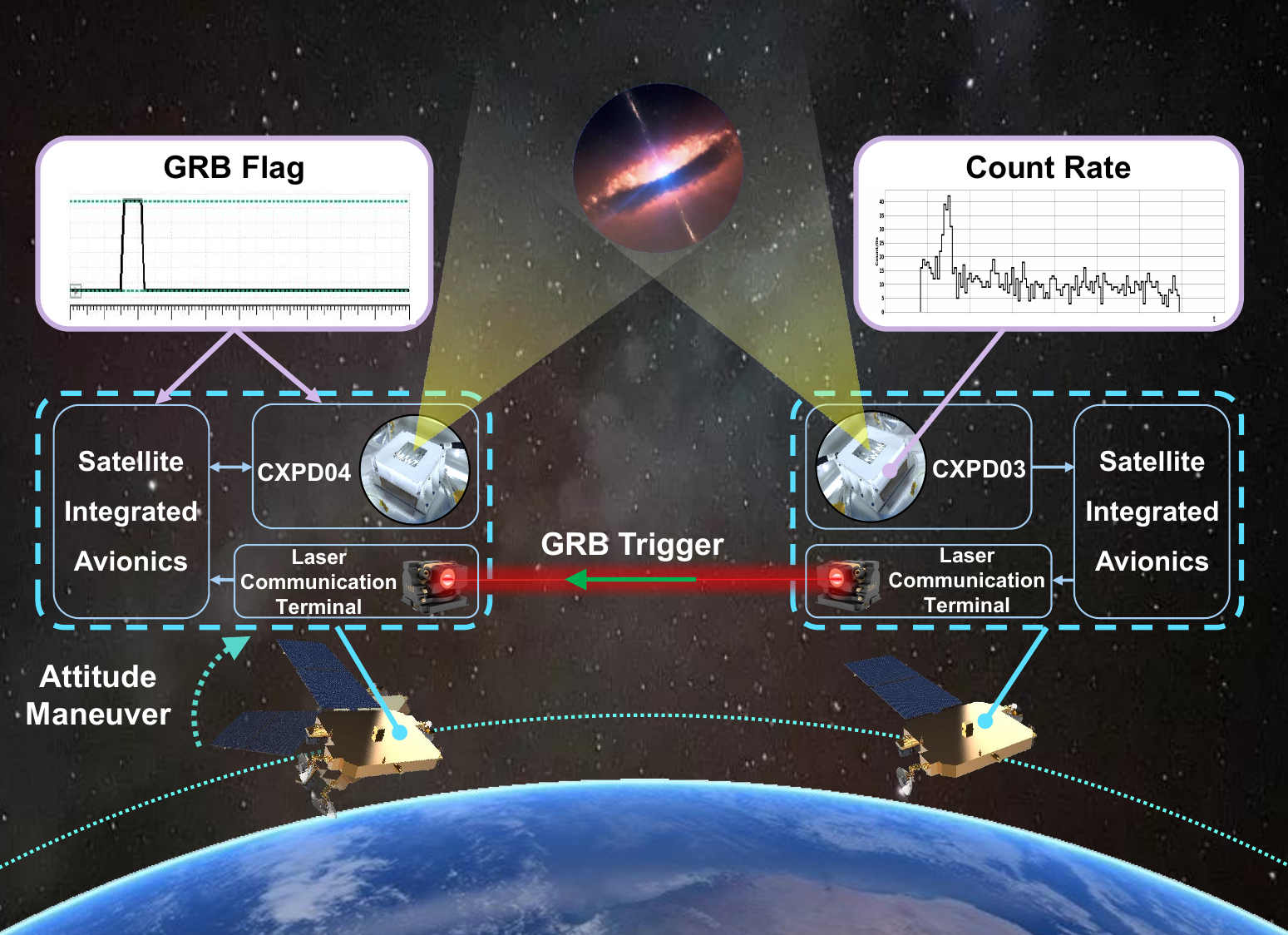}}
    \caption{Schematic diagram of autonomous dual-satellite coordinated observation control. Upon GRB detection, the primary satellite transmits a collaboration flag and attitude data to the companion via laser link. The companion verifies payload status and pointing constraints before commanding its AOCS for coordinated slew.}
    \label{fig:10}
\end{figure}

\subsection{GRB Trigger and Dual-Satellite Coordination}

The multi-satellite cooperative mode has the potential to significantly enhance the detection capability for random transient sources such as GRBs. Distributed satellites cover separate sky regions, enabling wider continuous monitoring and increasing the chance of capturing random transients. When one satellite initially detects a target, it can send a trigger via inter-satellite communication link to direct others to rapidly adjust attitude, allowing simultaneous joint observation. This coordination enhances both event statistics and multi-angle data for studies like polarimetry. GRBs typically last on the order of minutes, offering a narrow observing window. To validate this approach, the CXPD-Duo payload configuration employs a dual-satellite scheme, implementing an FPGA-based real-time GRB trigger Signal-to-Noise Ratio (SNR) algorithm\cite{ref32} defined as:

\begin{equation}
\label{eq:snr}
\mathrm{SNR} = \frac{ X_n - \sum\limits_{i=n-16}^{n-1} \frac{X_i}{16} }
                   { \sqrt{ \sum\limits_{i=n-16}^{n-1} \frac{X_i}{16} } }
                   > GRB\_Thr
\end{equation}

where $X_n$ is the event count in the current time unit and $\sum_{i=n-16}^{n-1} \frac{X_i}{16}$ is the average over the previous 16 units. In quiescent conditions, the event rate remains stable at about 2 counts/s, giving an SNR near zero\cite{ref19,ref20}. Simulations using soft X-ray GRB models indicate that when a GRB enters the field of view, the count rate can increase rapidly. To capture such events, the trigger threshold GRB\_Thr is set to 2, which can be updated in-orbit based on background conditions. To reduce FPGA resource use, the algorithm is implemented in a transformed form without explicit square-root or division operations:

\begin{equation}
\left[ \left( 16X_n - \sum_{i=n-16}^{n-1} X_i \right)_+ \right]^2 > 16GRB\_Thr^2 \sum_{i=n-16}^{n-1} X_i
\end{equation}

where $(16X_n-\sum_{i=n-16}^{n-1} X_i)_+$ denotes the positive-only function that ensures the numerator remains non-negative. To further reduce false triggers, the algorithm adds a minimum event-rate threshold and a duration check for sustained threshold exceedance, effectively suppressing background noise and weak GMCP discharges.

By processing and evaluating detector signals in real time, the algorithm generates GRB trigger commands that serve as key inputs for dual-satellite coordination. As shown in Fig.~\ref{fig:10}, a typical autonomous cooperative sequence operates as follows: When the primary satellite (e.g., CXPD-03) detects an event rate above threshold for over 5 s, it sends a trigger flag to SIA\_A. SIA\_A transmits the flag and pointing data via laser link to SIA\_B on the cooperating satellite. SIA\_B forwards the flag to CXPD-04 via CAN bus. CXPD-04 verifies that it is in observation mode and has not itself triggered; if confirmed, it returns a “payload ready for attitude adjustment” status. The cooperating SIA then assesses maneuver feasibility, ensuring the Sun does not enter the field of view and Earth does not block the new pointing. If conditions are met, attitude adjustment begins. The entire inter-satellite communication and status check completes within 5 s. One of the two payloads is slewed from its idle anti-Earth pointing to align with the other, requiring $\leq$ 1 min. Given limited ground validation, the cooperative mode will be confirmed on-orbit via the laser link after launch and refined through in-orbit debugging and firmware updates. The feasibility of intersatellite laser links has been demonstrated in the Space Computing Constellation, which provides an infrastructure for dual-satellite coordination\cite{ref33}. Ground testing focuses on verifying GRB trigger flag generation and communication protocols to establish the foundation for in-orbit operation.

\begin{figure}[t]
    \centering
    \begin{subfigure}[b]{0.46\textwidth}
        \centering
        \includegraphics[width=0.78\textwidth]{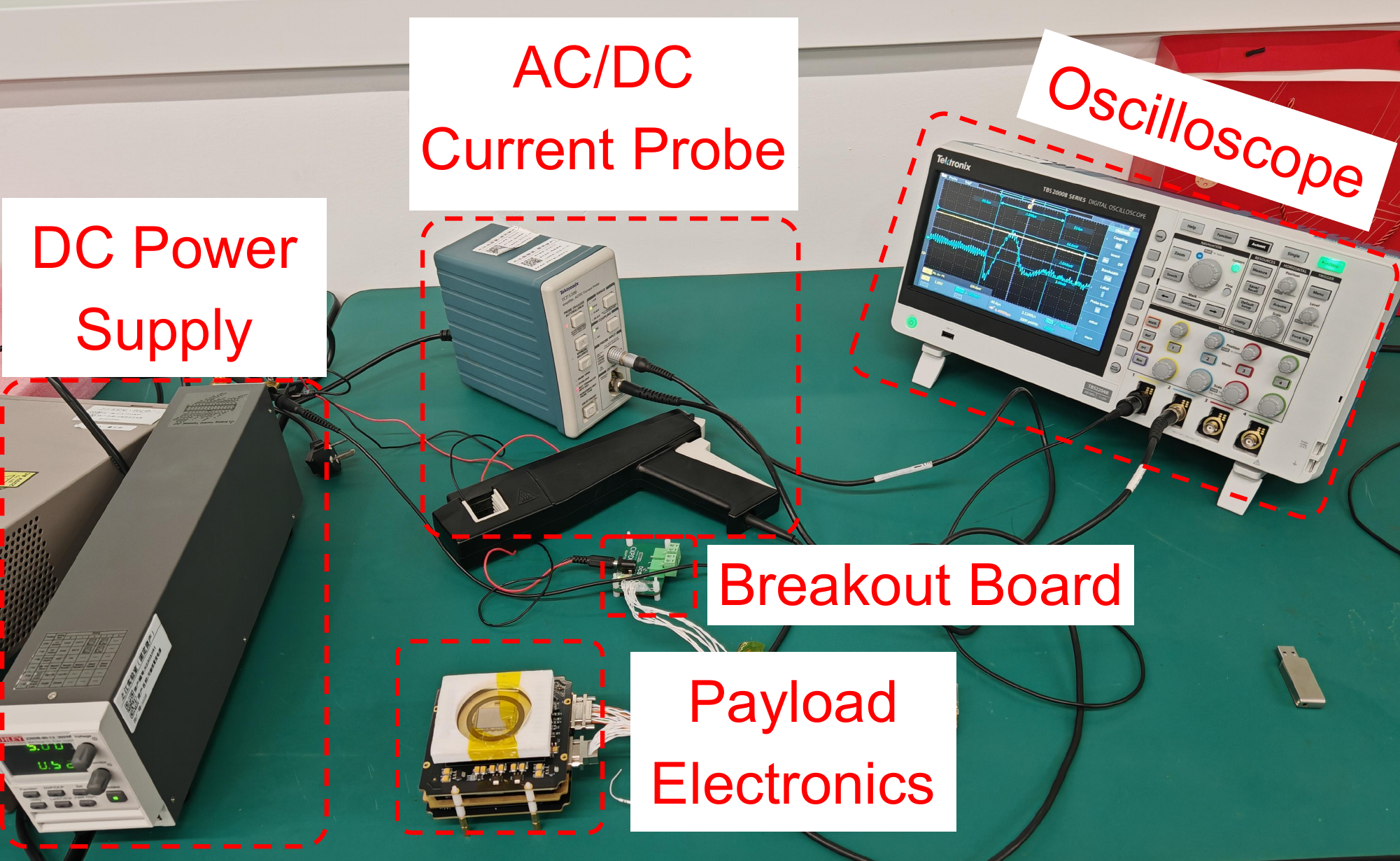}
        \caption{Test configuration including the payload electronics, current probe (set to 5 A/V), and oscilloscope.}
    \end{subfigure}
    \hfill
    \begin{subfigure}[b]{0.46\textwidth}
        \centering
        \includegraphics[width=0.78\textwidth]{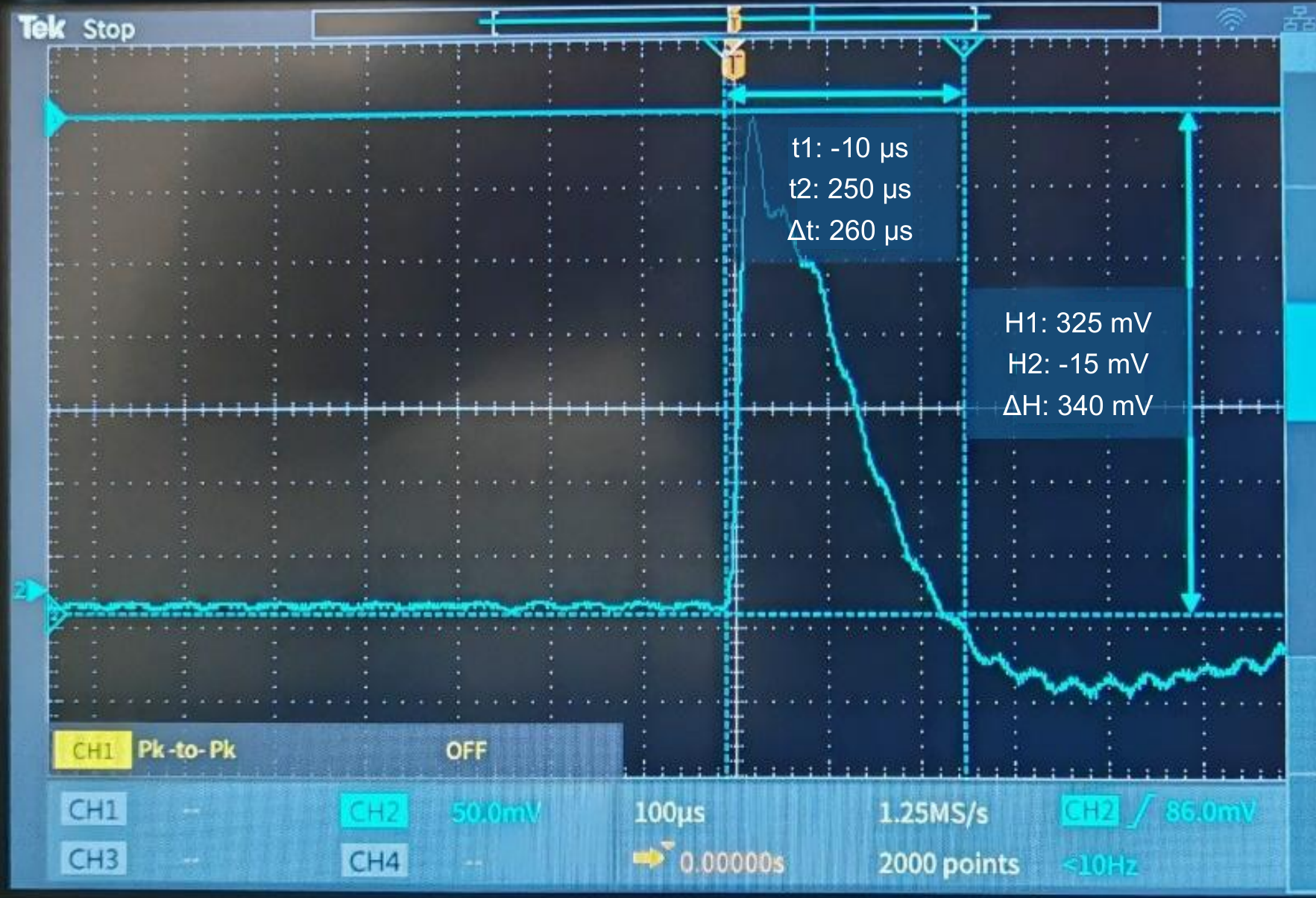}
        \caption{Power-on surge waveform at the 5 V input, showing a peak current of approximately 1.7 A and a duration of about 260 $\mu$s.}
    \end{subfigure}
    \caption{ Inrush current measurement setup and results.}
    \label{fig:11}
\end{figure}

\begin{figure}[t]
    \centering
    \begin{subfigure}[b]{0.46\textwidth}
        \centering
        \includegraphics[width=0.9\textwidth]{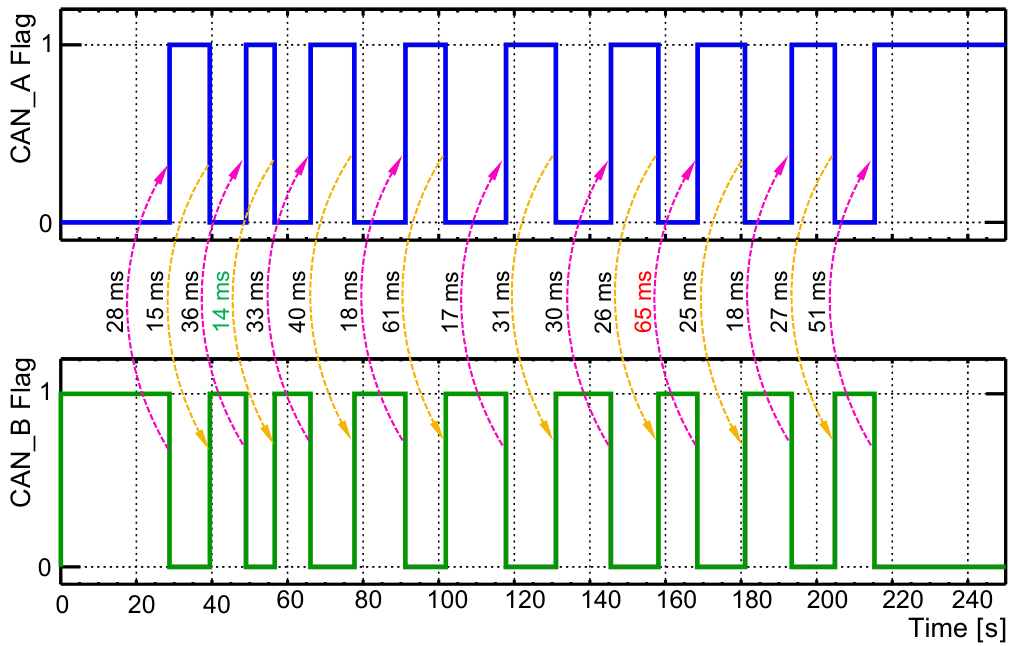}
        \caption{CAN bus channel switching waveform: a high level indicates the corresponding channel is active; the curved arrows mark the time interval between the falling and rising edges, representing the channel switching time.}
    \end{subfigure}
    \hfill
    \begin{subfigure}[b]{0.46\textwidth}
        \centering
        \includegraphics[width=0.9\textwidth]{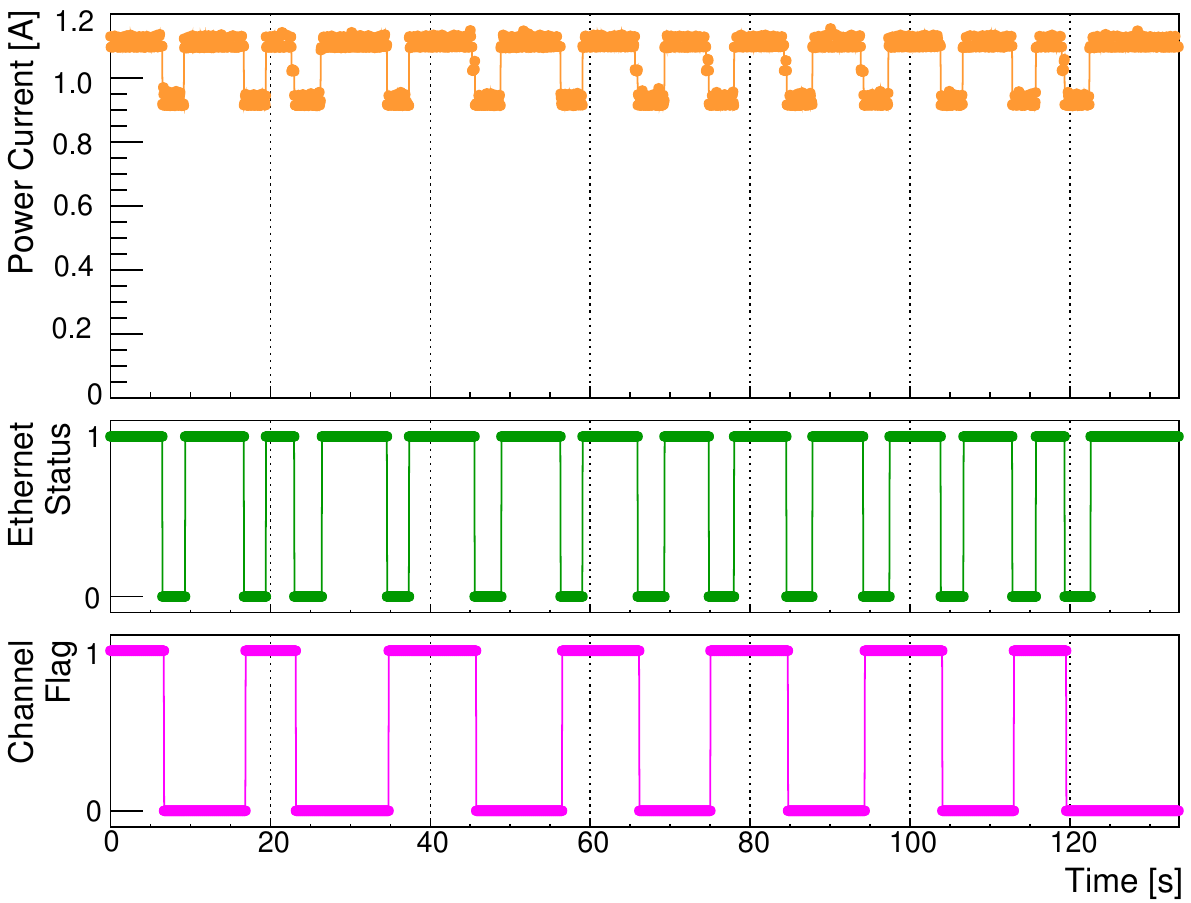}
        \caption{Ethernet channel switching waveform, from top to bottom: system total current, Ethernet link status (low = disconnected, high = link established), and channel flag (low = ETH\_A, high = ETH\_B).}
    \end{subfigure}
    \caption{Communication interface redundancy switching test results.}
    \label{fig:12}
\end{figure}

\begin{figure}[t]
\centerline{\includegraphics[width=3.28in]{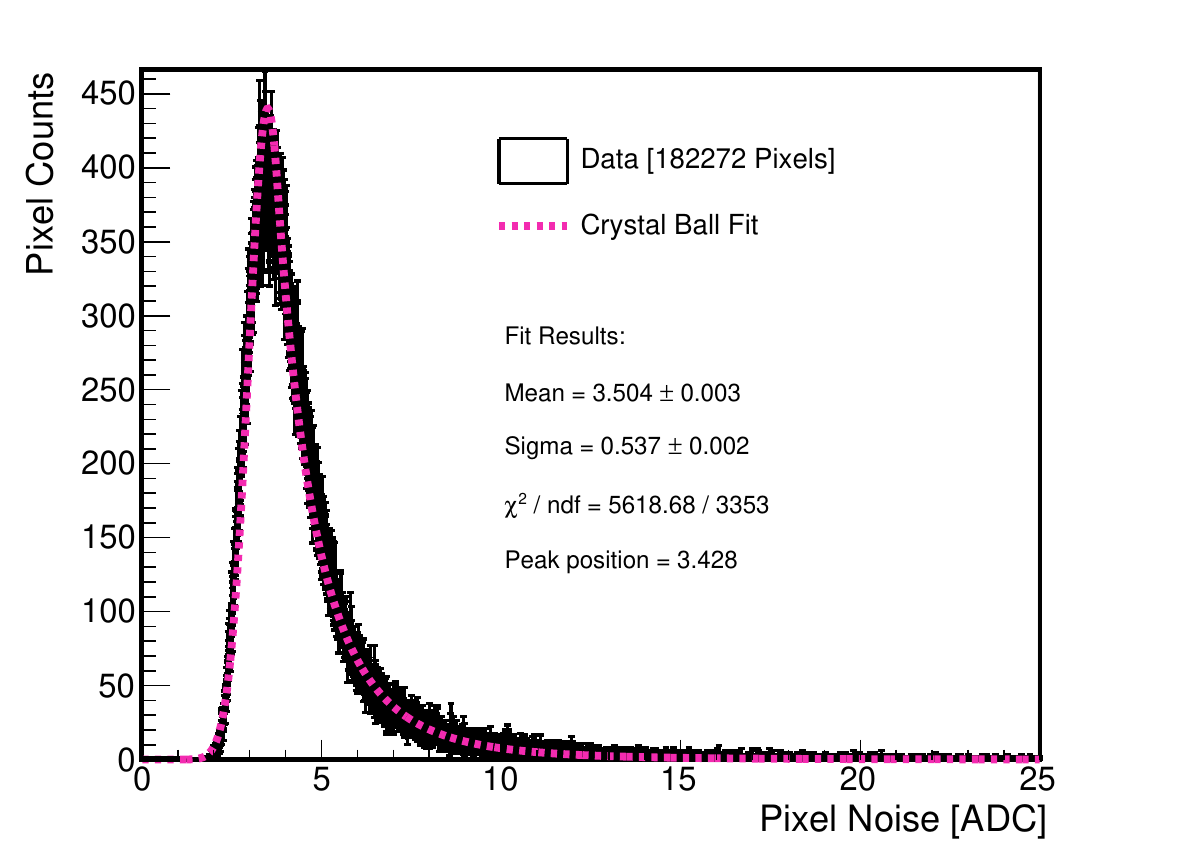}}
\caption{Background noise distribution of the Topmetal-L readout channels. A total of 182,272 pixels were sampled for output noise statistics.}
\label{fig:13}
\end{figure}

\begin{figure}[t]
\centerline{\includegraphics[width=2.8in]{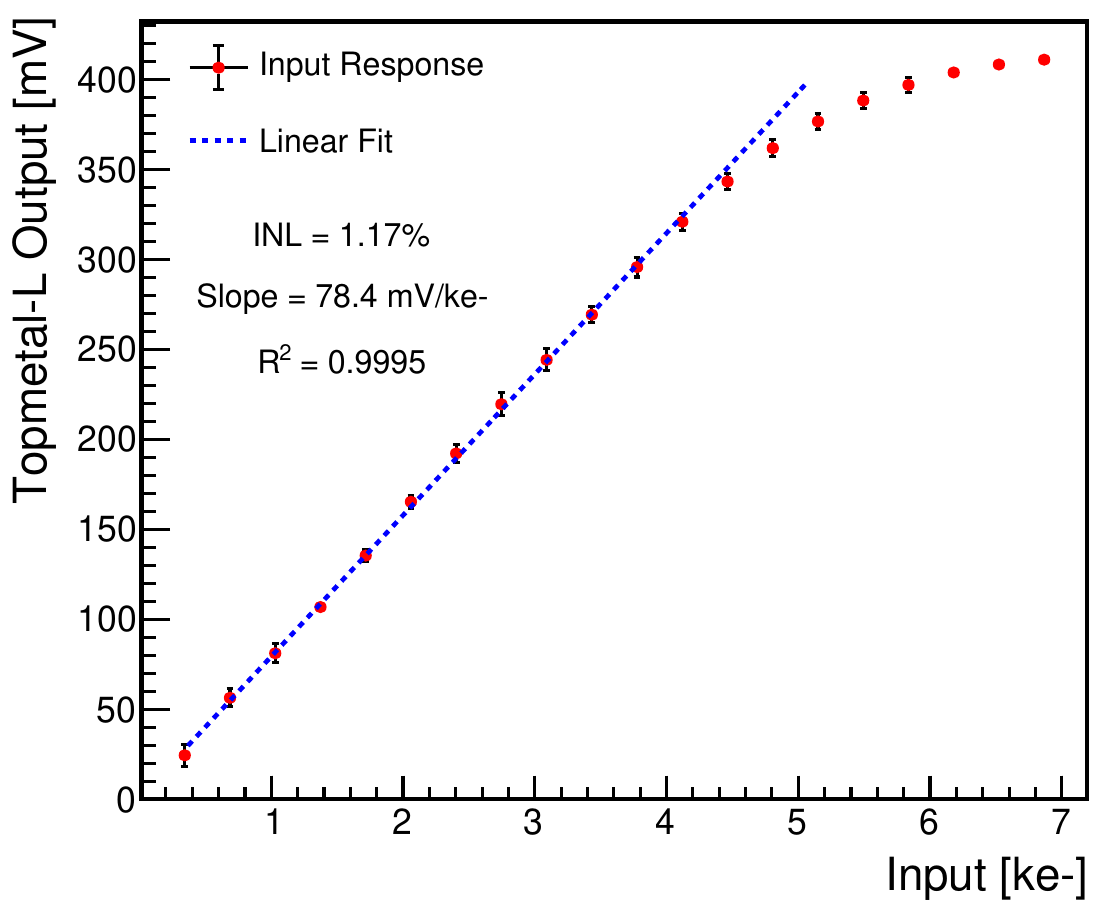}}
\caption{Topmetal-L channel input-output dynamic range and linearity of the readout channels.}
\label{fig:14}
\end{figure}

\begin{figure}[t]
\centerline{\includegraphics[width=3.38in]{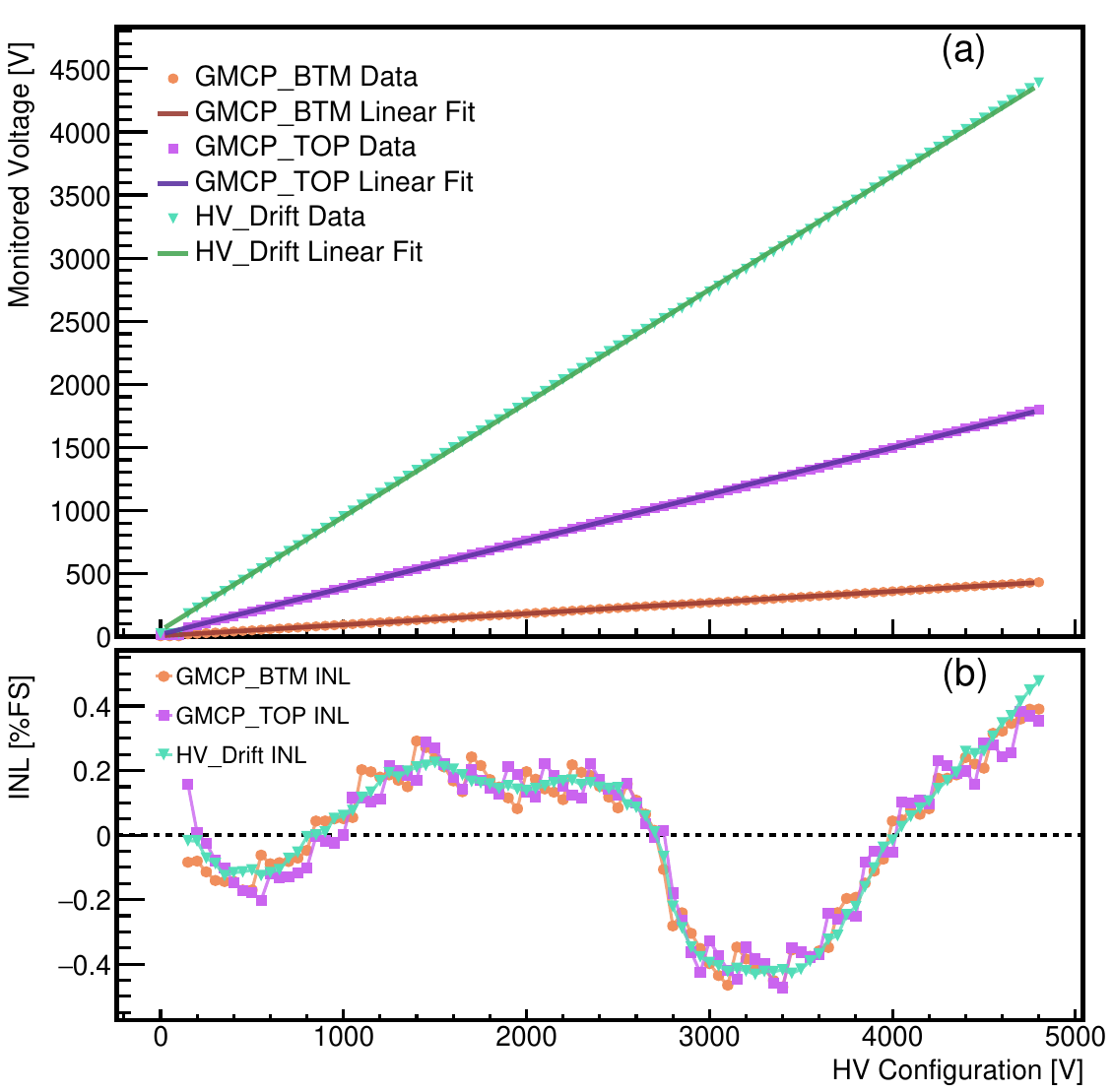}}
\caption{HV module output linearity test results. (a) Relationship curve between the system-measured feedback voltage and the set voltage; actual negative HV signals are displayed as absolute values. (b) Integral nonlinearity of the relationship curve.}
\label{fig:15}
\end{figure}

\section{Tests and Validation}
\label{sec:5}
\subsection{Power Supply and Communication Interface Tests}

\begin{figure*}[t]
\centerline{\includegraphics[width=6in]{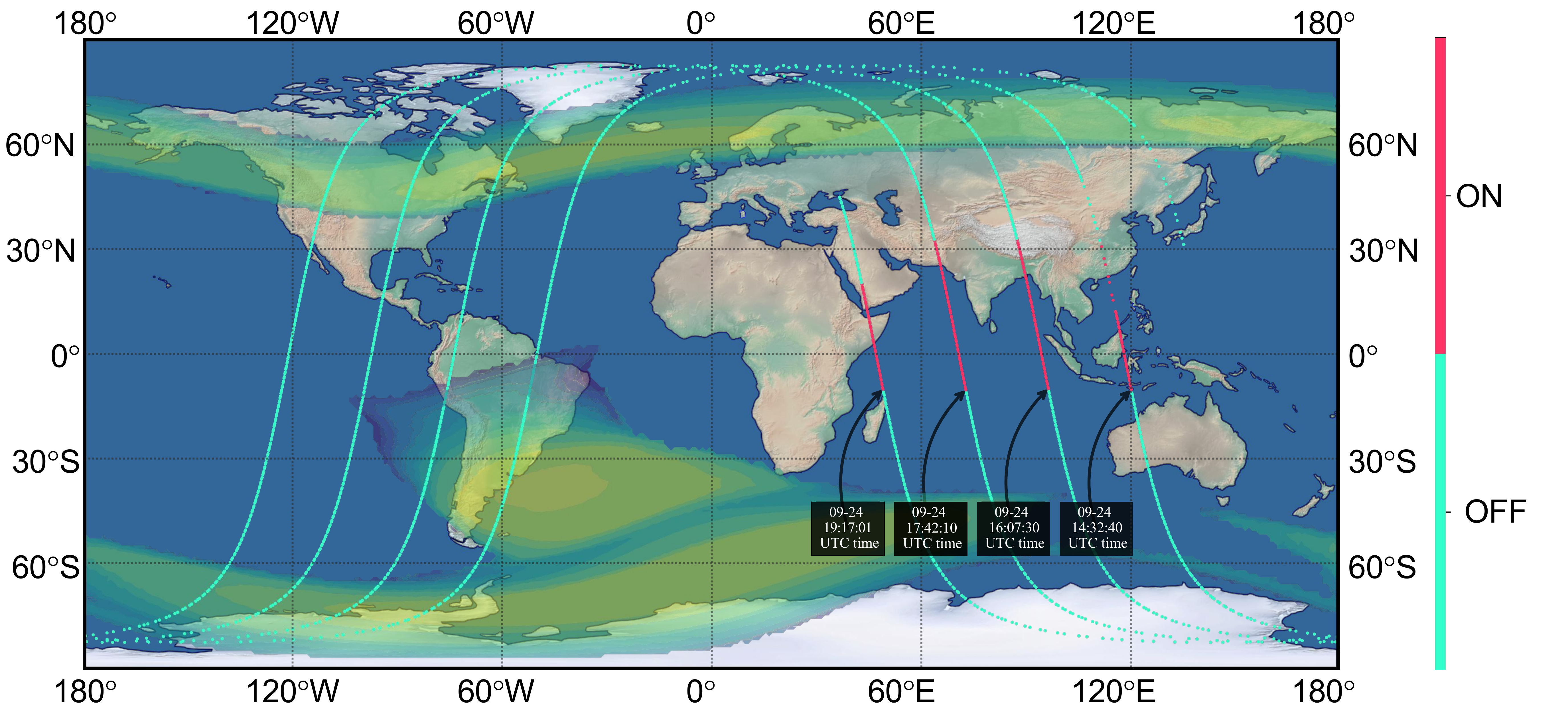}}
\caption{Ground simulation test of on-orbit autonomous HV management. The test simulated a strictly anti-Earth pointing attitude, with the system automatically controlling high voltage and data acquisition based on its operational state.}
\label{fig:16}
\end{figure*}

The CXPD-Duo system operates at a total power consumption below 6 W, supplied by the satellite bus's 5 V DC rail and converted via DC-DC and LDO modules. To prevent inrush current from interfering with shared equipment, a staged power-on sequence is employed: the core system activates first, followed by higher-power components. The Ethernet module is enabled only after FPGA configuration, and the Topmetal-L sensor remains in a low-power analog standby mode ($<$10 mW) until entering observation mode. Inrush characteristics were evaluated using the setup in Fig.~\ref{fig:11}(a), which includes a DC supply, current probe, oscilloscope, payload electronics, and a J30J adapter. With the current probe set to 5 A/V, the transient current during power-up was captured (Fig.~\ref{fig:11}(b)). The measured inrush current peaked at approximately 1.7 A with a duration of about 260 $\mu$s, meeting the power-supply constraints of the satellite bus.

To ensure reliable dual-satellite coordination, the payload communication interfaces are designed with a warm-standby redundant architecture, enabling primary-to-backup switching for both CAN and Ethernet channels. Fig.~\ref{fig:12}(a) presents the CAN channel switching test, during which 17 switching cycles were recorded using two SPDT switches per differential pair. Transition times ranged from 14 ms (minimum) to 65 ms (maximum), averaging about 32 ms, including switch actuation, with all transitions remaining under 100 ms. In the event that a CAN frame transmission times out due to the channel switching, the CAN controller’s hardware automatically retransmits the unacknowledged frame. With the CAN telemetry cycle operating at 1 s in coordination with SIA, this switching performance ensures reliable command exchange during dual-satellite operations. 

For the Ethernet data-downlink channel, switching speed requirements are more relaxed. As shown in Fig.~\ref{fig:12}(b), the system operates by default in gigabit Ethernet mode, drawing approximately 150 mA per active channel. A noticeable drop in total supply current occurs when the Ethernet link is disconnected. Across 13 switching cycles, the average link-establishment time was about 3.06 s, including physical switching overhead. In actual observation missions, the onboard router typically remains in standby to conserve power; the Ethernet link is activated only during high-throughput operations such as data downlink or firmware updates.

\subsection{Detector Performance Tests}

The performance of the Topmetal-L sensor was characterized using the CXPD-Duo electronics. The combined noise distribution of the Topmetal-L sensor and the readout electronics is shown in Fig.~\ref{fig:13}. Statistics collected from all pixels yield a distribution with a mean of 3.504 ADC and a standard deviation of 0.537 ADC. The voltage conversion factor of the readout chain is approximately 0.5 mV/ADC. The measured mean noise of 3.504 ADC, therefore, corresponds to an average noise voltage of about 1.752 mV. 

\begin{figure}[t]
\centerline{\includegraphics[width=3.38in]{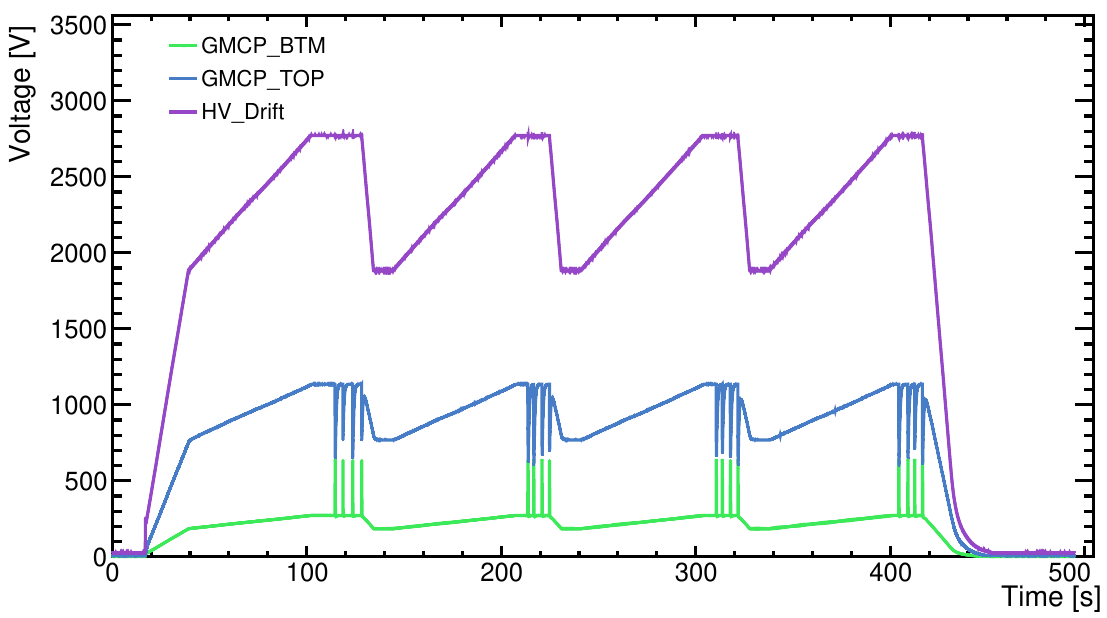}}
\caption{Operation of the HV discharge protection mechanism. The test illustrates controlled ramp-up and the protection sequence: after multiple discharges, the voltage reduces to Safe\_Vol and recovers if no further events occur; repeated triggers within 12 min initiate an anomaly shutdown.}
\label{fig:17}
\end{figure}

The full-channel gain was measured by injecting a negative step signal through the chip’s test input. Dynamic-range test results are presented in Fig.~\ref{fig:14}. In the input range of 0--4 ke$^{-}$, the pixel channel exhibits a well-behaved linear response; above 4 ke$^{-}$, the response deviates from linearity and gradually saturates. Fitting the linear region yields a charge-to-voltage conversion gain of approximately 78.4 mV/ke$^{-}$, with a coefficient of determination R$^{2}$ = 0.9995 and an integral nonlinearity of 1.17\%, indicating excellent linearity within the effective dynamic range. From the gain and noise measurements, the total equivalent noise charge of the full channel is calculated as 22.35 e$^{-}$.

\subsection{HV System Function Tests}

The HV module supports continuous adjustment of its output up to -5 kV via a 0--5 V analog control signal but does not include built-in output-voltage monitoring. The relationship between the monitored output voltage and the input control voltage is shown in Fig.~\ref{fig:15}. When the absolute value of the control voltage is below 200 V, the HV module produces no output. Above 200 V, all three monitoring branches exhibit a linear response with strong consistency among channels; integral nonlinearity is better than $\pm$0.5\%. In actual observation missions, full-scale output is typically unnecessary, so the system enforces an upper limit on the control voltage, preventing further increase beyond 4800 V.

CXPD-Duo verified its on-orbit autonomous control through a closed-loop orbital simulation. The simulation employed actual sun-synchronous orbital parameters while the detector maintained a strict anti-Earth pointing attitude throughout. Due to the CXPD-Duo’s wide field of view, the Sun enters the detector’s field when the satellite is on the sunlit side. With a simulated orbital period of 1.5 h, the system autonomously enabled high voltage and started data acquisition when entering the observable dark side, and disabled both when entering the SAA or sunlit side. The observable region was defined conservatively for initial operational safety. Fig.~\ref{fig:16} shows the control timing over four consecutive orbits, with a total duration of 6 h. High voltage and observation mode were consistently activated during each dark-side interval, while the system remained in a safe standby state otherwise. All monitoring functions remained active, with operational status recorded in real time in the eMMC for subsequent on-orbit health diagnostics and mission analysis.

To ensure GMPD safety, the system implements controlled HV ramp-up/ramp-down and discharge protection. The sequence is shown in Fig.~\ref{fig:17}, with Safe\_Vol set to 2000 V and the output target to 3000 V. During ramp-up, the rate is 100 V/s below Safe\_Vol and 20 V/s above it. Discharge protection uses two thresholds (Discharge\_Thr = Protect\_Thr = 3). If four normal discharges occur briefly, output drops to Safe\_Vol; if no discharge follows within 10 s, voltage is restored at 20 V/s. If protection triggers three times in 12 min, the fourth triggers an HV anomaly shutdown. All shutdowns use a fast 200 V/s ramp-down to reach a safe state promptly.

\begin{figure}[t]
\centerline{\includegraphics[width=3.08in]{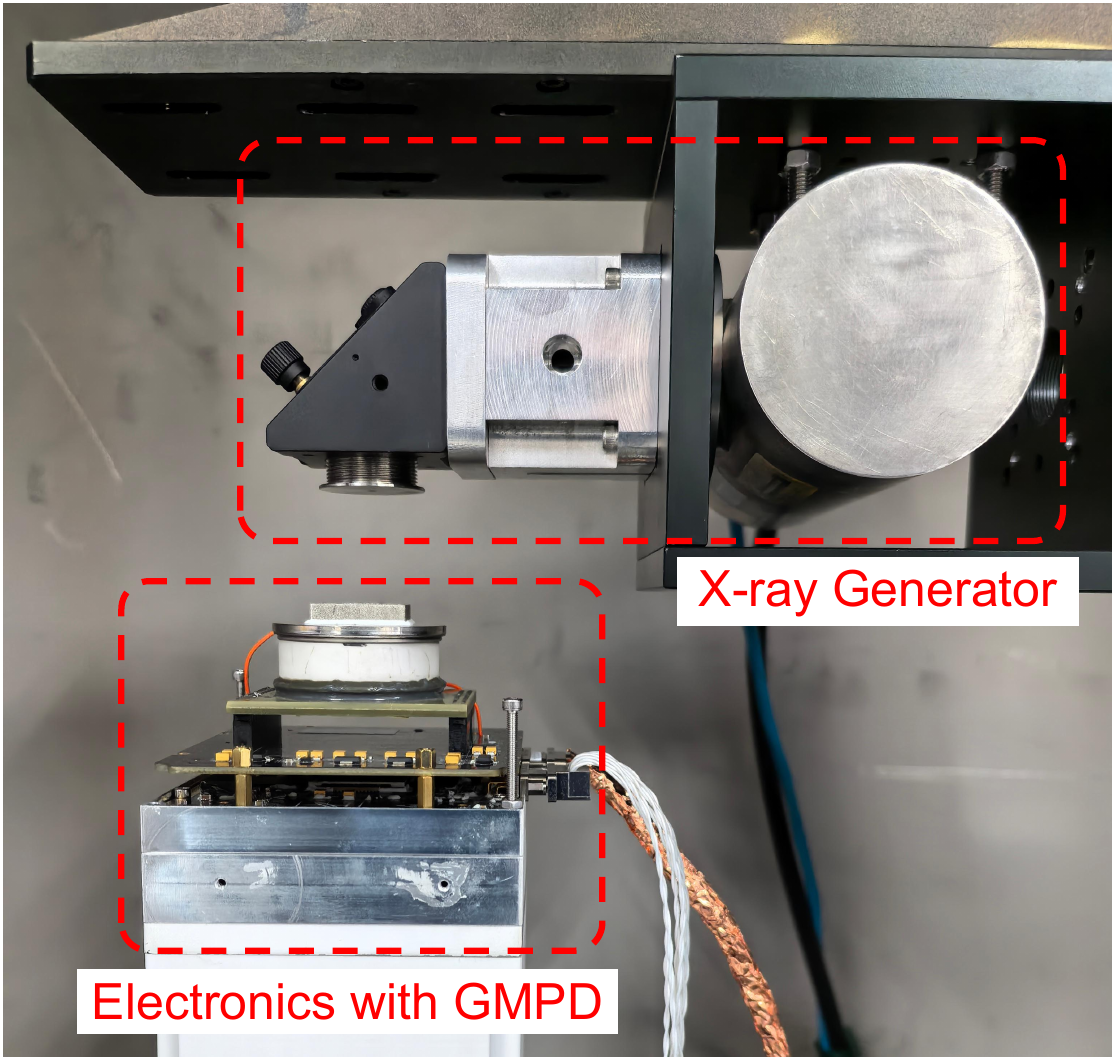}}
\caption{X-ray test system for the GMPD. The X-ray generator’s current is adjustable to control flux, with the source collimated and aligned to the GMPD entrance window.}
\label{fig:18}
\end{figure}

\begin{figure}[t]
\centerline{\includegraphics[width=3.38in]{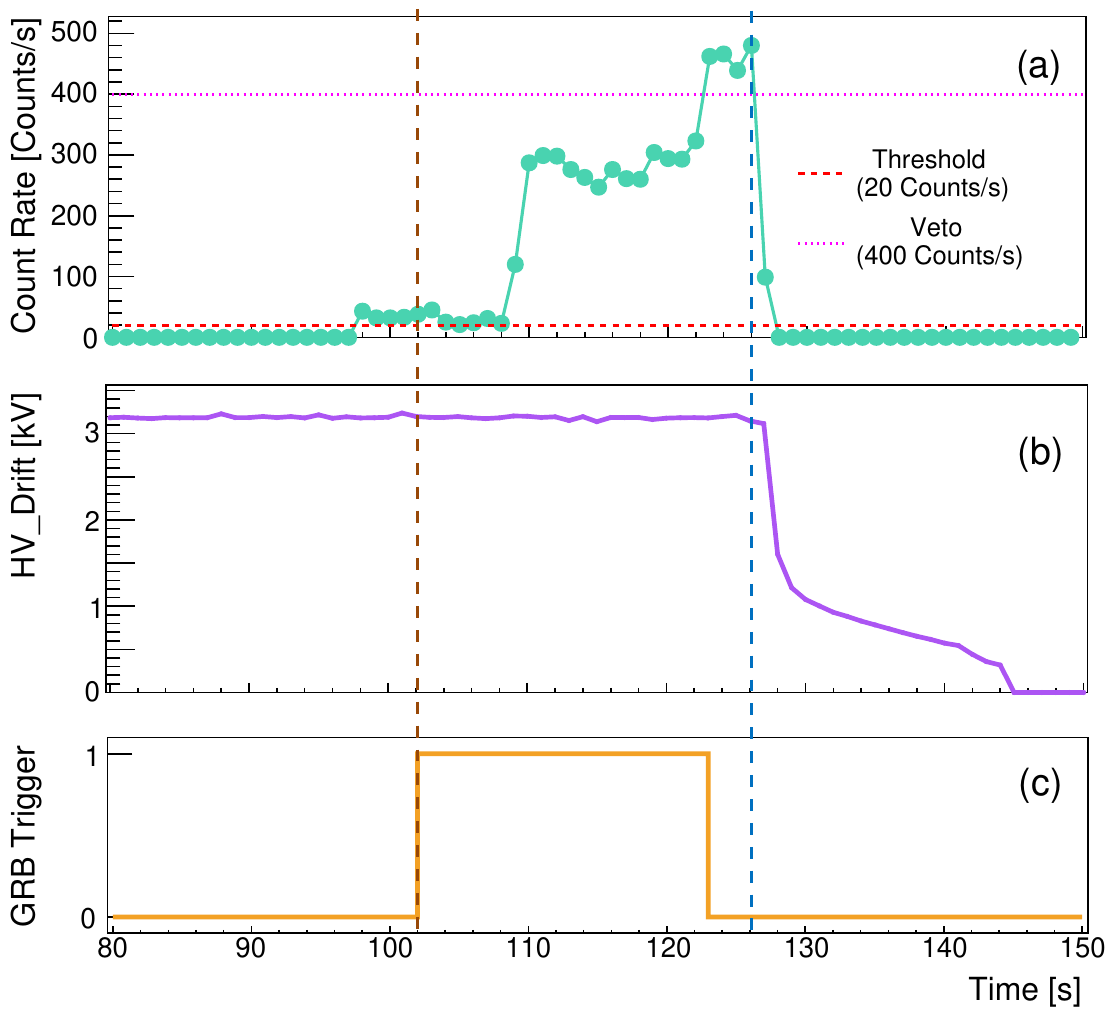}}
\caption{System behavior test under different count rates, showing the delayed response of the GRB trigger flag after the X-ray source is enabled, and the HV anomaly shutdown triggered when the count rate becomes excessive. (a) Detector count rate; (b) HV\_Drift monitoring; (c) GRB trigger flag.}
\label{fig:19}
\end{figure}

\begin{figure}[t]
    \centering
    \begin{subfigure}[b]{0.35\textwidth}
        \centering
        \includegraphics[width=\textwidth]{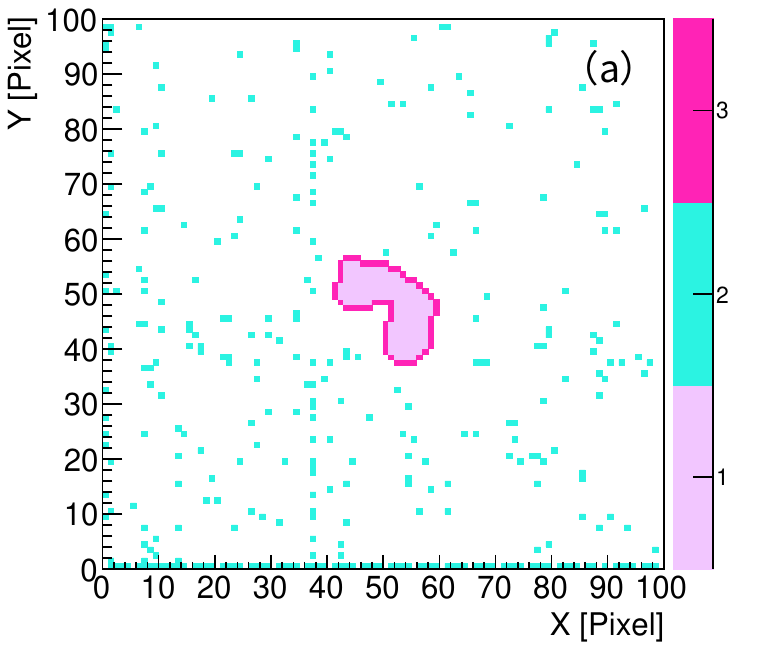}
        \caption{ }
    \end{subfigure}
    \hfill
    \begin{subfigure}[b]{0.36\textwidth}
        \centering
        \includegraphics[width=\textwidth]{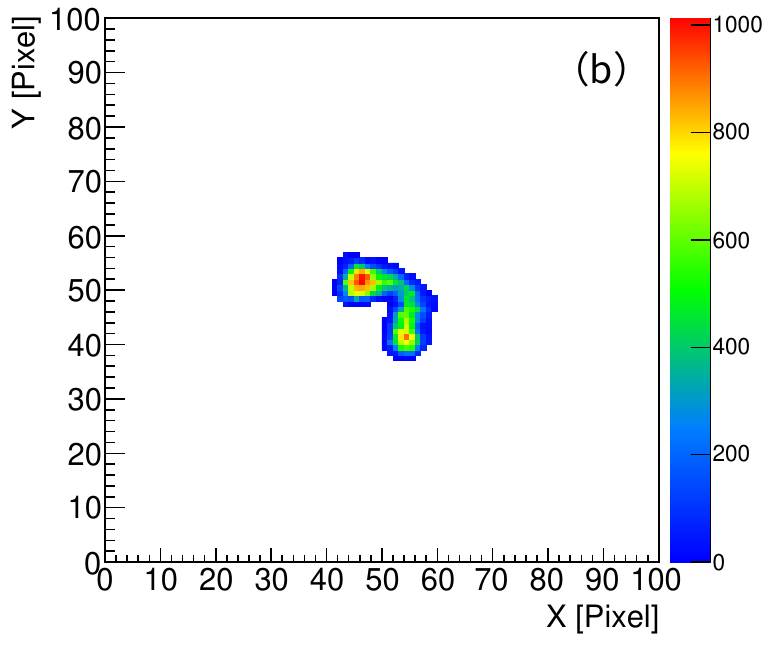}
        \caption{ }
    \end{subfigure}
    \caption{Real-time compression workflow and results of GMPD photoelectron tracks. (a) Schematic of binarization and morphological opening (erosion followed by dilation): 1 denotes pixels belonging to the binarized track core, 2 represents above-threshold pixels removed as noise, and 3 indicates track-edge pixels restored after dilation. (b) Final photoelectron track image after the above processing.}
    \label{fig:20}
\end{figure}

To address potential damage from abnormal increases in detector count rates, the HV protection system was enhanced with a fast, count-rate-based protection mechanism in addition to the existing discharge protection. This mechanism is designed to prevent the automatic HV control routine from erroneously enabling high voltage under abnormal conditions, such as when the satellite traverses SAA or when the Sun enters the field of view. The test setup is shown in Fig.~\ref{fig:18}. An adjustable-flux X-ray generator, collimated and aligned with the GMPD detector, was used to simulate different radiation environments by varying its operating current.

The test procedure and results are presented in Fig.~\ref{fig:19}. First, the detector was configured in its on-orbit observation state, with the high voltage raised to an absolute working value of 3450 V. The count-rate threshold for HV anomaly protection was set to 400 counts/s for this test, and it can be adjusted in-orbit via command from 0 to 65535 counts/s in steps of 1. Without the X-ray source, the detector count rate in the ground-based background remained near 0 counts/s. After the source was turned on, the count rate increased and stabilized at approximately 20 counts/s. After remaining above this level for more than 5 seconds, the system correctly triggered and maintained the GRB flag. Gradually increasing the X-ray flux raised the count rate to 250--300 counts/s, and the detector continued to operate normally. Further increasing the radiation flux caused the count rate to exceed 400 counts/s, at which point the GRB flag was cleared. If the count rate remained above this threshold for 5 consecutive seconds, the HV anomaly protection routine was triggered: the high voltage was first rapidly reduced below Safe\_Vol, then ramped down to zero at a controlled rate to avoid voltage transients that could cause secondary stress on the detector.

\subsection{Track Data Compression Processing}

The Topmetal-L sensor, as a large-array pixel detector, supports user-defined readout regions. By integrating it with an ADC and an FPGA, this system implements region-of-interest readout: once triggered, only a 100 $\times$ 100 pixel region is read, thereby reducing invalid data at the source. To further alleviate resource constraints on data storage and downlink, real-time onboard track compression is applied before storage. Photoelectron tracks collected by the Topmetal-L in the GMPD appear as locally clustered signals, with useful information typically concentrated in limited pixel areas. Storing full-frame pixel images would introduce substantial background noise, occupying FPGA buffer and eMMC memory and saturating the downlink bandwidth. In extreme cases, buffer overflow could even cause the loss of valid scientific data.

To address this issue, a track-image compression algorithm based on binary morphological opening is employed. The algorithm first applies a threshold to the original image, generating a binary mask that separates track signals from background noise. However, simple thresholding still retains many isolated noise pixels and may degrade track-edge information. The algorithm, therefore, performs an opening operation on the binary result: an erosion step removes scattered single-pixel noise, followed by a dilation step that restores the morphology of continuous track regions. This process suppresses noise while preserving track-shape integrity. Fig.~\ref{fig:20} illustrates the full opening-operation sequence applied to a 5.9 keV photoelectron track using a 3 $\times$ 3 convolution kernel. By retaining sufficient information for track reconstruction and polarization analysis, this method significantly reduces the volume of background data to be stored and transmitted, effectively easing the onboard data-processing burden on the payload.

\subsection{Thermal Vacuum Cycling Test}

\begin{figure}[t]
\centerline{\includegraphics[width=3.38in]{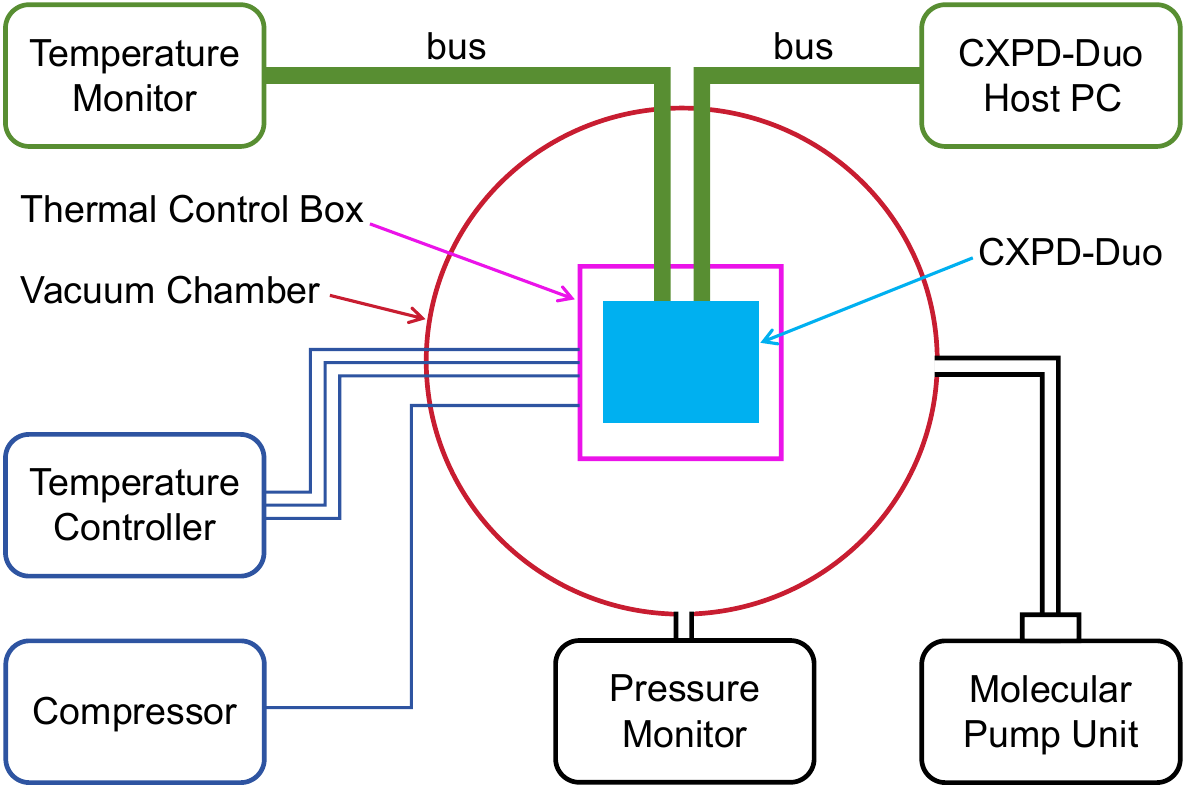}}
\caption{Schematic diagram of the thermal vacuum circulation system.}
\label{fig:21}
\end{figure}

\begin{figure}[t]
\centerline{\includegraphics[width=3.38in]{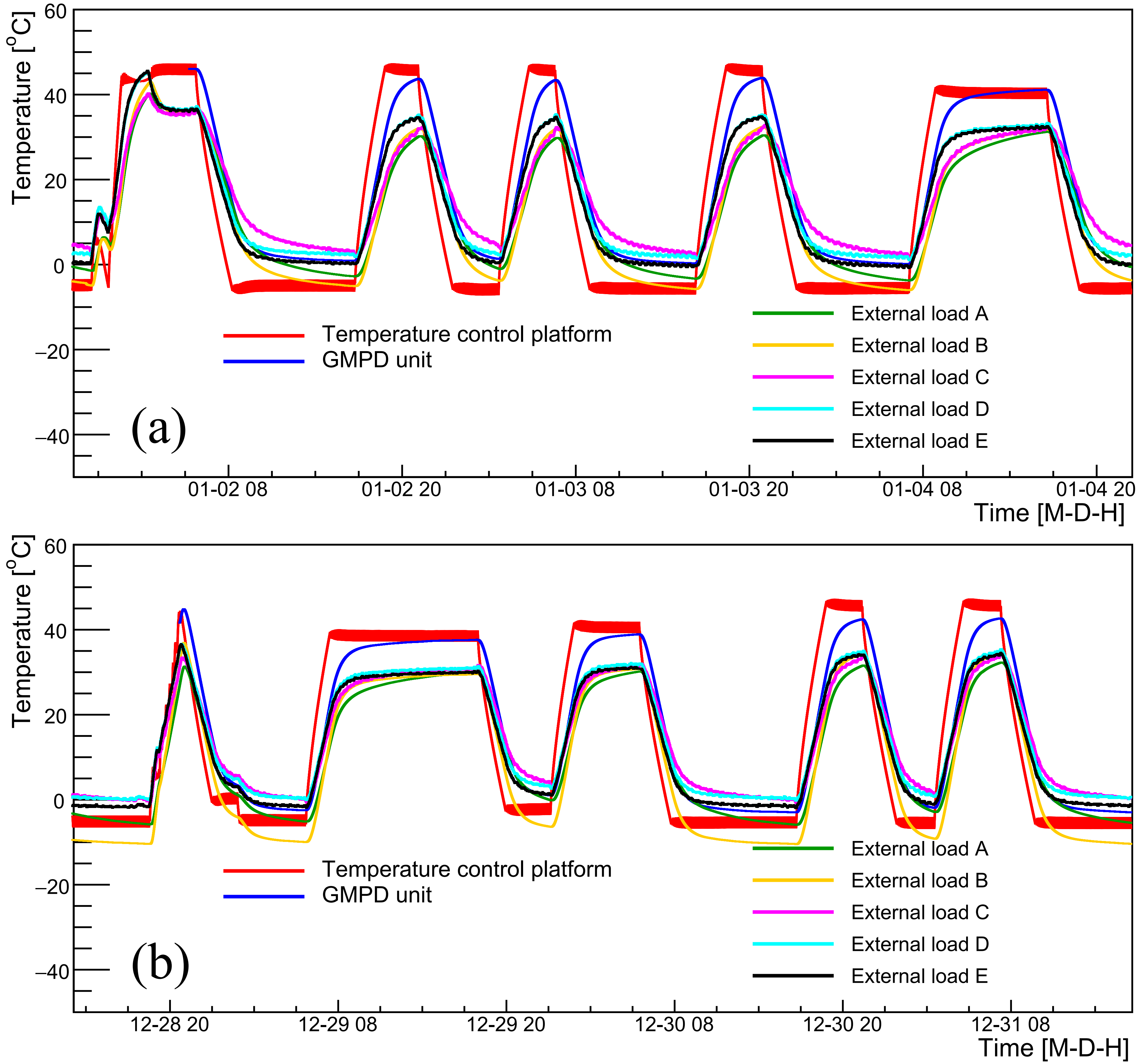}}
\caption{Temperature variations at various monitoring points during normal operation of CXPD-Duo. Panels (a) and (b) show the thermal cycling curves for the CXPD-03 and CXPD-04 payloads, respectively. Curves labeled External load A, B, C, D, and E correspond to different surfaces of the CXPD-Duo housing.}
\label{fig:22}
\end{figure}

To characterize the vacuum thermal performance of the CXPD payload, a thermal vacuum cycling system was established, comprising a vacuum chamber, a temperature controller, a compressor, a molecular pump unit, a pressure monitor, temperature monitors, and a host personal computer (PC), as shown in Fig.~\ref{fig:21}. Inside the system, a gold-plated aluminum alloy temperature-controlled box can maintain temperatures from -50 $^\circ$C to 52 $^\circ$C under vacuum. The vacuum chamber operates at a pressure of approximately 10$^{-4}$ Pa. This setup was used for thermal vacuum testing of the CXPD CubeSat payloads.

During the test, the CXPD payload was mounted inside the temperature-controlled box with its mounting surface in close contact with the bottom of the box. Temperature sensors were placed on several external surfaces. Under vacuum, thermal cycling tests were performed on two CXPD payloads while they were in operation. The temperature of the controlled box was cycled between -5 $^\circ$C and 40 $^\circ$C for a total of five cycles. The measured temperature variations at the monitoring points are presented in Fig.~\ref{fig:22}. Panels (a) and (b) show the thermal cycling curves for the CXPD-03 and CXPD-04 payloads, respectively. After the pressure inside the vacuum chamber had stabilized, HV tests lasting more than 30 minutes were performed during both high-temperature and low-temperature plateau phases. No abnormal discharge was observed; only normal discharges occurred at a rate of approximately two events per five minutes. The internal temperature was approximately 5--10 $^\circ$C higher than the external side temperature, with the maximum internal temperature reaching 45 $^\circ$C. Throughout the test, the payloads operated normally with no anomalies. After thermal cycling, the scientific performance of the payloads was re-evaluated, and the results were consistent with those measured before the test.

Additionally, to verify the detector’s tolerance to extreme low temperatures, the payload was powered off and the environmental temperature was lowered to -50$^\circ$C for 12 hours. After returning to room temperature, the detector operation and monitoring data remained normal.

\section{Conclusion}
\label{sec:6}

This study presents the design, implementation, and verification of the CXPD-Duo electronics system for dual-satellite coordinated X-ray polarimetry. Operating within stringent CubeSat constraints, the system integrates a GMPD and Topmetal-L sensor into a compact hardware platform that performs full-chain signal processing, onboard data reduction, and autonomous coordinated control. The core innovations address three principal challenges:  (1) designing an HV safety management mechanism with controlled ramp-up/ramp-down and dual protection against both discharges and count-rate anomalies; (2) establishing a coordinated control flow based on real-time triggering and redundant communication, validating the feasibility of autonomous dual-satellite cooperative observation; (3) implementing region-of-interest readout and morphological compression algorithms, which significantly alleviate storage and downlink burdens for track data.

Ground tests demonstrated an equivalent noise charge of 22.35 e$^{-}$, communication switching times under 100 ms, and compliance of all key performance metrics with mission requirements. Thermal vacuum cycling tests confirmed stable operation under simulated space conditions, with no functional degradation after five cycles between -5 $^\circ$C and 40 $^\circ$C. The sensitivity of a single payload is limited, but the dual-satellite cooperative architecture validated here provides a feasible pathway toward multi-unit constellations, which is expected to improve the minimum detectable polarization. This work not only delivers a viable engineering solution for space-science observation under limited resources, but also lays a technical foundation for future intelligent micro-/nano-satellite cooperative networks. Future work may leverage the system’s in-orbit reconfigurability and integrate the astronomical large model to further enhance the intelligence of data preprocessing and autonomous system management.

\vfill

\end{document}